\documentclass[fleqn,usenatbib]{mnras}

\usepackage{newtxtext}
\usepackage{newtxmath}
\usepackage[T1]{fontenc}
\usepackage{ae,aecompl}

\usepackage{graphicx}	% Including figure files
\newcommand{\lya}{Ly$\alpha$}
\newcommand{\HI}{\ion{H}{i}}
\newcommand{\DA}{D_A}
\newcommand{\nhi}{N_{\rm HI}}

\newcommand{\hmpc}{h^{-1}{\rm Mpc}}
\newcommand{\gHI}{\Gamma_{\rm HI}}
\newcommand{\kms}{\;{\rm km}\,{\rm s}^{-1}}
\newcommand{\persecond}{\;{\rm s}^{-1}}
\newcommand{\msolar}{\;{\rm M}_{\odot}}
\newcommand\cdunits{{\rm cm}^{-2}}

\newcommand{\mufasa}{{\sc Mufasa}}
\newcommand{\simba}{{\sc Simba}}

\newcommand{\mbh}{M_{\rm BH}}

\newcommand{\fgam}{F_{\rm UVB}}
\newcommand{\siglim}{\sigma_{\rm lim}}

\newcommand{\romeel}[1]{\color{black}{#1}\color{black}}
\newcommand{\jacob}[1]{\color{black}{#1}\color{black}}
\newcommand{\romeelnew}[1]{\color{black}{#1}\color{black}}
\newcommand{\jacobnew}[1]{\color{black}{#1}\color{black}}

\title[Jet Feedback and the PUC in \simba]{Jet Feedback and the Photon Underproduction Crisis in \simba} 

\author[Christiansen et al.]{
Jacob F. Christiansen$^{1}$\thanks{E-mail: jacobfc96@yahoo.com},
Romeel Dav\'e$^{1,2,3}$,
Daniele Sorini$^{1}$,
Daniel Angl\'es-Alc\'azar$^{4,5}$
\\
$^{1}$ Institute for Astronomy, Royal Observatory, Edinburgh EH9 3HJ, United Kingdom\\
$^{2}$ University of the Western Cape, Bellville, Cape Town 7535, South Africa\\
$^{3}$ South African Astronomical Observatories, Observatory, Cape Town 7925, South Africa\\
$^4$ Center for Computational Astrophysics, Flatiron Institute, 162 Fifth Avenue, New York, NY 10010, USA\\
$^5$ Department of Physics, University of Connecticut, 196 Auditorium Road, U-3046, Storrs, CT 06269-3046, USA\\
}

\date{Accepted XXX. Received YYY; in original form ZZZ}

\pubyear{2019}

\begin{document}
\label{firstpage}
\pagerange{\pageref{firstpage}--\pageref{lastpage}}
%%TC:endignore
\maketitle

\begin{abstract}
We examine the impact of black hole jet feedback on the properties of the low-redshift intergalactic medium (IGM) in the \simba\ simulation, with a focus on the \lya\ forest mean flux decrement $D_A$. Without jet feedback, we confirm the Photon Underproduction Crisis (PUC) in which $\gHI$ at $z=0$ must be increased by $\times6$ over the Haardt \& Madau value in order to match the observed $\DA$. Turning on jet feedback lowers this discrepancy to $\sim\times 2.5$, and additionally using the recent Faucher-Gigu{\`e}re \romeelnew{background mostly resolves the PUC, along with producing a flux probability distribution function in accord with observations.}\  The PUC becomes apparent at late epochs ($z\la 1$) where the jet and no-jet simulations diverge; at higher redshifts \simba\ reproduces the observed $\DA$ with no adjustment, with or without jets.  The main impact of jet feedback is to lower the cosmic baryon fraction in the diffuse IGM from 39\% to 16\% at $z=0$, while increasing the warm-hot intergalactic medium (WHIM) baryon fraction from 30\% to 70\%; the lowering of the diffuse IGM content directly translates into a lowering of $\DA$ by a similar factor.  Comparing to the older \mufasa\ simulation that employs different quenching feedback but is otherwise similar to \simba, \mufasa\ matches $\DA$ less well than \simba, suggesting that low-redshift measurements of $\DA$ and $\gHI$ could provide constraints on feedback mechanisms.  Our results suggest that widespread IGM heating at late times is a plausible solution to the PUC, and that \simba's jet AGN feedback model, included to quench massive galaxies, approximately yields this required heating.
\end{abstract}
%%TC:ignore
\begin{keywords}
methods: numerical; galaxies: evolution; galaxies: formation; galaxies: intergalactic medium; quasars: absorption lines
\end{keywords}

%%%%%%%%%%%%%%%%%%%%%%%%%%%%%%%%%%%%%%%%%%%%%%%%%%

%%%%%%%%%%%%%%%%% BODY OF PAPER %%%%%%%%%%%%%%%%%%

\section{Introduction}
\label{sec:intro}

\subsection{Background}

The intergalactic medium (IGM) contains the vast majority of cosmic baryons at all cosmic epochs~\citep{meiksin-2009}.  After the epoch of reionisation, the IGM is highly ionised by a cosmic background of ultraviolet photons (UVB) emitted by star forming galaxies and active galactic nuclei (AGN).  The trace neutral component is detectable as \HI\ \lya\ absorption in the spectra of background sources such as quasars, which is known as the Lyman alpha forest.  The temperature of this gas is set by a balance between local adiabatic expansion and photo-heating from the metagalactic flux, leading to a relatively simple equation of state~\citep{hui-gnedin-1997}.  Combined with the fact that absorbing gas mostly tracks gravitationally-driven large-scale structure, this has made the \lya\ forest useful for a wide range of cosmological applications.

The optical depth $\tau$ of Lyman alpha forest absorbing gas along a given line of sight (LOS) depends on the gas density and the neutral fraction.  The neutral fraction is itself proportional to the density and inversely proportional to the \HI\ photo-ionisation rate ($\gHI$).  If we consider the mean optical depth in the \lya\ forest, it thus scales as the square of the mean baryonic density (which is $\propto \Omega_b$), and inversely with $\gHI$: 
$\bar\tau \propto \Omega_b^2/\gHI$, with constants that depend on cosmology~\citep{rauch-1997}, and a small correction owing to the temperature dependence of the \HI\ recombination rate.
The fluctuations around this mean optical depth can thus be used to measure the matter power spectrum, assuming that the baryons trace matter~\citep[e.g.][]{weinberg-1998}.  The mean optical depth, meanwhile, can be used to constrain a combination of $\Omega_b$ and $\gHI$.

\citet{rauch-1997} applied this approach to measurements of the mean flux decrement in the \lya\ forest at $z\sim 2-3$ in order to estimate $\Omega_b$, assuming $\gHI$ taken from  \citet{haardt-madau-1996}, and obtained $\Omega_b>0.021h^2$.  The \citet{haardt-madau-1996} background was estimated from the number density of observed quasars and star-forming galaxies plus radiative transfer through a clumpy IGM, assuming that all ionising photons from quasars and a small fraction of such photons from star-forming galaxies escaped. Despite substantial uncertainties in source count observations at that time, this value for $\Omega_b$ turned out to be in good agreement with determinations from the deuterium abundance~\citep{tytler-1996} and subsequently the cosmic microwave background~\citep{planck-2016}.

At lower redshifts, the growth of the Cosmic Web results in gas shock-heating on filamentary structures as it accretes supersonically~\citep{dave-1999}.  This generates the so-called Warm Hot Intergalactic Medium~\citep[WHIM;][]{cen-ostriker-1999,dave-2001a} of gas outside bound halos in the $T\sim 10^5-10^7$K temperature range.  Owing to the nonlinear processes involved, gas dynamical simulations are required to study the growth of the WHIM, and concomitantly, the reduction in \lya\ forest baryons.  Such simulations broadly predict that roughly one-third of cosmic baryons at the present epoch are in the WHIM~\citep{dave-2001a,dave-2010, smith-2011}.  It is very challenging to detect such warm-hot gas observationally since the hydrogen is fully ionised, so metal line absorbers must be used instead, which are weaker and more uncertain.  Nonetheless, an observational census primarily from \ion{O}{vi} absorption suggests that such predictions are broadly consistent with current data~\citep{tripp-2001,shull-2012}.

In spite of the increased complexity introduced by the WHIM, it is still possible to use the \lya\ forest mean flux decrement to measure $\gHI$, given that $\Omega_b$ is now well-determined from other avenues. Indeed, at $z\sim 0$, this is currently the most robust approach to measuring $\gHI$, because it is impossible to directly detect the 912\AA\ photon background directly given foreground Galactic absorption, and other approaches such as H$\alpha$ fluorescence are extremely challenging~\citep[though see][]{fumagalli-2017}.  \citet{dave-2001b} used this approach on {\it Hubble} Space Telescope Imaging Spectrograph data to measure $\gHI(z=0)=10^{-13.3\pm0.7}\persecond$.  In the meantime, \citet{haardt-madau-2001} had improved upon their estimate of $\gHI$ evolution from source count modeling, and determined $\gHI(z=0)=10^{-13.08}\persecond$, consistent with the \lya\ forest measurements.  Thus it appeared that $\gHI$ at $z=0$ was now pinned down to within a factor of a couple.

Measurements of cosmic ionising photon sources continued to improve.  In particular, it became clear that the assumption in \citet{haardt-madau-2001} of a constant 10\% escape fraction of Lyman continuum photons from galaxies was inconsistent with observations; stacked measures of dwarf galaxies at intermediate redshifts suggested instead values below 2\%~\citep{rutkowski-2016}.  \citet{faucher-2009} did a new calculation of $\gHI(z)$, and estimated $\gHI(z=0)=3.9\times10^{-14}\persecond$.  \citet{haardt-madau-2012} further updated their estimate assuming an evolving escape fraction of $1.8\times 10^{-4}(1+z)^{3.4}$ and found an even lower $\gHI(z=0)=2.3\times 10^{-14}\persecond$. Hence as these calculations became more precise, they diverged substantially from the original determination by \citet{haardt-madau-2001} of $\gHI(z=0)=8.3\times 10^{-14}\persecond$, with the latest determinations lower by nearly a factor of four.

In light of this, \citet{kollmeier-2014} re-investigated constraints on $\gHI$ at $z=0$ from the \lya\ forest using new simulations that were substantially improved in dynamic range and input physics compared to those in \citet{dave-2001a}.  This study was also enabled by an improved census of \lya\ forest absorbers from {\it Hubble}'s Cosmic Origins Spectrograph (COS) by \citet{danforth-2016}.  \citet{kollmeier-2014} found that, in order to match the amplitude of the observed column density distribution or the mean flux decrement, it was necessary to increase the \citet[hereafter HM12]{haardt-madau-2012} value of $\gHI(z=0)$ by a factor of $\approx 5$, i.e. $\gHI(z=0)\approx 10^{-13}\persecond$.  In other words, if the \lya\ forest is robustly predicted in simulations as expected from the simple physics involved, then there was a gross shortfall of observed photon sources relative to that needed to match the observed IGM ionisation level.  Most of the newfound discrepancy owed to the change in the source count estimates of $\gHI$.  What had initially seemed like a solved problem in 2001 was now, with improved measurements and simulations, yielding a substantial discrepancy. \citet{kollmeier-2014} dubbed this the Photon Underproduction Crisis (PUC) -- the Universe did not seem to be producing nearly enough photons to explain the ionisation level seen in the \lya\ forest.

\subsection{Previous investigations into the PUC}

\romeel{The PUC could potentially be solved in a number of ways:  (i) it could be that the ionising background strength in HM12 was underestimated; (ii) it could be that the simulations are simply incorrect, due to numerics; (iii) it could be that new physics impacts the diffuse IGM, causing it to be more ionised.  The subsequent results have been somewhat disparate and controversial, but as we argue below, it is becoming clear that the PUC indeed exists at a level comparable to that presented in \citet{kollmeier-2014}.

There has been growing consensus that the HM12 ionising background may be too weak at $z=0$.  Potential systematic uncertainties include not only the source population emissivity and escape fraction, along with the \lya\ column density distribution at the high-$N_{HI}$ end which provides a sink term for ionising photons.  The ionising background model of \citet{faucher-2009} predicts $\gHI(z=0)$ about twice the HM12 value, which would partially mitigate the PUC.  \citet{khaire-2015} did a UV background calculation using updated QSO emissivities that were $2\times$ higher than those in \citet{haardt-madau-2012}, and suggested that this combined with a 4\% escape fraction from galaxies could increase the source count estimate of $\gHI$ up to the levels required to match \citet{kollmeier-2014}.  While their assumed QSO emissivity is plausible, the 4\% global escape fraction of ionising photons from galaxies seems less plausible given current measurements \citep[e.g.][]{rutkowski-2016}.  A recent update of the \citet{faucher-2009} background in \citet[hereafter FG19]{faucher-2019} similarly found a value of $\gHI(z=0)$ about twice that in \citet{haardt-madau-2012}.  Furthermore, a recent determination of $\gHI(z)$ from \citet{khaire-2019} preferred a higher value for $\gHI(z=0)$, but not by more than a factor of two. \citet{kulkarni-2019} found that AGN can only account for half the required photons even though they are expected to greatly dominate the low-$z$ ionising photon budget. Hence it appears that a $\sim\times 2$ systematic difference on the determination of $\gHI$ in HM12 is reasonable.  However, a factor of $\sim 5$ seems difficult to accommodate, as no recent background yields such a large difference (since \citealt{haardt-madau-2001}).  Thus the PUC may yet reflect some underlying missing physics in the low-$z$ IGM.

The second option is to appeal to numerics in order to ionise the IGM and lower the required flux.  Certainly, the complicated non-linear growth of structure driving the WHIM may be subject to the details of hydrodynamical or other methodology.  In general, theoretical studies of the PUC studies can be divided into two classes: Those using cosmological hydrodynamical simulations including galaxy formation physics, and those that do not.  We recap results from the latter category first.}

\citet{shull-2015} compared measurements of the mean flux decrement from COS versus uniform-mesh {\sc Enzo} simulations, and determined $\gHI(z=0)=4.6 \times 10^{-14}\persecond$.  While still a factor of two off from the \citet{haardt-madau-2012} value, this could be probably accommodated within systematic uncertainties in $\gHI$.  However, there are two significant caveats.  First, uniform-mesh simulations are known to overproduce entropy in low-Mach number shocks and hence increase the amount of numerical heating in the IGM; indeed, in the \citet{dave-2001a} comparison of the WHIM in various simulations, the fixed mesh code of \citet{cen-ostriker-1999} yielded $\ga 50\%$ the baryons in the WHIM, while adaptive resolution codes (both Eulerian and Lagrangian) yielded $\sim 30\%$.  Second, their predicted \lya\ absorber column density distribution was substantially steeper than observed, so while at high column densities ($\nhi\sim 10^{14}\cdunits$) the amplitude agreed with \citet{haardt-madau-2012}, at low columns ($\nhi\sim 10^{13}\cdunits$), it agreed better with \citet{haardt-madau-2001}.  

\romeel{\citet{gaikwad-2017a} and \citet{gaikwad-2017b} employed a model based on non-radiative hydrodynamic simulations, post-processed to mock up the impact of shock heating and radiative cooling.  They also found a substantially reduced PUC, about a factor of $\sim 2$, which they argued could be accommodated by a change in the photo-ionising background.  While their model enabled efficient exploration of parameter space and could be calibrated to match simulation results in some diagnostics, it is unclear that such a post-processed treatment of heating and cooling can accurately capture the highly non-equilibrium thermodynamics in the IGM.

\citet{viel-2017} examined the PUC in two hydrodynamics simulations:  Sherwood~\citep{bolton-2017} and Illustris~\citep{vogelsberger-2014}. Sherwood has a large volume but low resolution and did not include star formation or feedback.  Illustris included these, and particularly strong AGN feedback.  \citet{viel-2017} concluded that in both cases, there was a preference for $\gHI(z=0)$ being $\sim\times 1.5-3$ higher than the HM12 value. They further highlighted \lya\ linewidths as a potential discriminant between models.

In contrast to these studies, the results from high-resolution simulations with well-constrained galaxy formation physics paint a different picture.  These include the original \citet{kollmeier-2014} result, but also studies with different numerical techniques.  \citet{tonnesen-2017} confirmed the \citet{kollmeier-2014} result using adaptive mesh refinement simulations with {\sc Enzo}, which suggested that the PUC is not sensitive to hydrodynamics methodology.  Results from \citet{gurvich-2017} with Illustris using the {\sc Arepo} moving mesh code also showed a similar PUC if no AGN feedback is included; we discuss these results further below.  Hence the numerics of the hydrodynamic solver does not seem to play a role, so long as one has high resolution and full galaxy formation physics. 
}

If the solution to the PUC cannot be fully solved obtained by appealing to uncertainties in source population modeling or numerical methodology, then the remaining potential solution is that models are missing some widespread IGM heating mechanism that would lower the \lya\ absorbing gas.  \citet{kollmeier-2014} investigated whether the then-popular blazar heating model of \citet{broderick-2012} could accommodate this, and determined that it could go partway, but it produced a column density distribution that was shallower than observed.  Since the mean flux tends to be dominated by near-saturated lines ($\nhi\sim 10^{13.7}\cdunits$) occurring in mildly over-dense regions~\citep{dave-1999}, it was not possible to solve the PUC by mostly heating void gas.  \citet{wakker-2015} strengthened the case for the PUC using the same simulations as \citet{kollmeier-2014} but a different observational measure, by showing that the \ion{H}{I} column density as a function of filament impact parameter required the HM12 ionizing background at $z=0$ to be increased by $\times 4-5$.

\romeel{A landmark study was that of \citet{gurvich-2017}, who investigated the PUC in Illustris.}\ Unlike most previous simulations studying the PUC, Illustris included strong AGN feedback.  This was primarily designed to quench star formation in massive galaxies by heating halo gas, but as a by-product it also deposited energy \romeel{not only in the vicinity of the AGN, hence decreasing \lya\ absorption in quasar environs \citep{sorini-2018, Sorini-phd}, but also into the diffuse IGM gas, thus affecting the \lya\ forest. Specifically, by comparing the fiducial Illustris run to a run with no AGN feedback, \citet{gurvich-2017} showed that such injection of energy into the diffuse IGM clearly went towards resolving the PUC in Illustris.}\ Assuming a~\citet{faucher-2009} UVB ($\sim\times 2$ higher than HM12), \citet{gurvich-2017} was able to match the observed mean flux decrement, although their column density distribution slope did not match COS data. Such a large impact from feedback was somewhat surprising, since it is commonly believed that galactic feedback does not strongly impact the diffuse IGM far from galaxies. 
%A benchmark study was that of \citet{gurvich-2017}, who investigated the PUC in Illustris.  Unlike most previous simulations studying the PUC, Illustris included strong AGN feedback.  This was primarily designed to quench star formation in massive galaxies by heating halo gas, but as a by-product it also deposited energy in the diffuse IGM.  \romeel{By comparing to a run with no AGN feedback, they showed that the latter effect clearly went towards resolving the PUC in Illustris.}\ Assuming a~\citet{faucher-2009} UVB ($\sim\times 2$ higher than HM12), \citet{gurvich-2017} was able to match the observed mean flux decrement, although their column density distribution slope did not match COS data. Such a large impact from feedback was somewhat surprising, since it is commonly believed that galactic feedback does not strongly impact the diffuse IGM far from galaxies.  

Although a promising solution to the PUC, Illustris at the same time greatly over-evacuates gas from massive halos~\citep{genel-2014}, so it is likely that their AGN feedback model is too strong, or adds energy in the wrong manner.  \romeel{These results also clarify the findings of \citet{viel-2017}, which at face value suggested that the inclusion of AGN feedback did not have a large impact owing to the similarity of results between Sherwood and Illustris. \citet{gurvich-2017} instead showed that this was not the case: AGN feedback had a substantial effect, and the agreement owed to comparing a low resolution simulation with no galaxy formation to one with full galaxy formation physics but including AGN.

In short, hydrodynamic models that include full galaxy formation physics constrained to match a variety of other observations all uniformly show a PUC at the $\sim\times 4-6$ level.  It is possible that $\times 2$ of this may be explained via a reevaluation of the low-redshift photo-ionisation rate from HM12.  However, this still leaves a factor of $\sim 2-3$ to explain.  Surprisingly, AGN feedback may impact the diffuse IGM far from galaxies to mitigate the PUC, but it remains unclear whether a fully successful model can be developed that self-consistently reproduces both galaxies and the low-$z$ \lya\ forest.  This is the goal of our present study.}

\subsection{This work: Exploring AGN feedback in the IGM with \simba}

AGN feedback is crucial for reproducing the observed galaxy population~\citep{somerville-2015}, but it is not well understood.  Recent years have seen the development of various AGN feedback models within cosmological hydrodynamic simulations, primarily designed to quench massive galaxies as observed~\citep{vogelsberger-2014,schaye-2015,weinberger-2018,henden-2018}.  One successful recent model is the \simba\ simulation. \simba\ uses an observationally-motivated two-mode feedback model, where at high Eddington rates it follows observed ionised or molecular gas outflow scalings, while at low Edington rates it switches to a jet mode with outflow speeds up to $\sim8000 \kms$, \romeel{broadly implementing the physical scenario outlined in \citet{best-2012}.}\ The two-mode approach is qualitatively similar to the model in IllustrisTNG~\citep{weinberger-2018}, although \simba\ uses stably bipolar outflows and significantly less total energy which is more consistent with observations of the kinetic  power in radio jets~\citep[e.g.][]{whittam-2018}.  Such jet feedback can potentially carry matter and hence energy far away from its host galaxy into the diffuse IGM~\citep{borrow-2019}.  \simba\ is able to quench galaxies in good agreement with observations over cosmic time, and more relevantly for this work, yields a hot baryon fraction in massive halos that is consistent with observations~\citep{dave-2019}, so is not over- or under-evacuating halo baryons.  \romeel{Moreover, it reproduces observed X-ray scaling relations in groups and clusters \citep{robson-2020}.}\ Hence it provides a plausible AGN feedback model that can be used to investigate the PUC.

In this paper we examine the PUC in the \simba\ simulation.  To do so, we generate simulated lines of sight in \lya\ absorption, and quantify the variation needed in the strength of the assumed photo-ionising background in order to match observations of the mean flux decrement $D_A$.  We focus on $D_A$ and not the column density distribution of absorbers in order to avoid uncertainties associated with line identification and fitting, which can be quite sensitive to spectral resolution and signal to noise~\citep[e.g.][]{dave-2001a}. In particular, we investigate the role of the jet mode of AGN feedback in \simba.  We show that this type of AGN feedback has a large impact on the PUC, while other AGN feedback modes in \simba\ (cf. radiative and X-ray) have minimal impact.  We also compare to the \mufasa\ simulation results, which assumed a different halo-based quenching model that did not employ jets, though still matched massive galaxy properties.  We find that \simba's AGN jet feedback model is crucial for obtaining agreement between the $\gHI$ required to match the $D_A$ observations and modern determinations of $\gHI$ from source population modeling, suggesting that widespread IGM heating from AGN is a key factor in helping to solve the PUC.

This paper is organised as follows. In \S\ref{sec:sims} we review the \simba\ simulations used in this work.  In \S\ref{sec:igm-background} we present some global IGM physical characteristics in the \simba\ runs with and without AGN jets. In \S\ref{sec:PUC} we present our main results in examining the PUC in \simba\ in runs with and without jets.  In \S\ref{sec:variants} we discuss the PUC in other AGN feedback tests in \simba, and in \mufasa. \jacob{In \S\ref{sec:modelling-uncertainties} we discuss various modelling uncertainties.}\ In \S\ref{sec:summary} we summarise our results.  

\section{The \simba\ Simulations}  \label{sec:sims}

\subsection{Input physics and cosmology}

\simba~\citep{dave-2019} is a cosmological hydrodynamic simulation that uses a Meshless Finite Mass (MFM) hydrodynamics solver~\citep{hopkins-2015}, which can be classified as an Arbitrary Lagrangian Eulerian (ALE) code.  MFM employs a Riemann solver that is able to handle strong shocks and shear flows accurately, without introducing an artificial viscosity~\citep{hopkins-2015}.  This is particularly beneficial in situations where high Mach number flows and strong shocks are an important physical aspect in the problem, which is the case here in studying the impact of high-velocity jet outflows (described below) on diffuse IGM gas.

\simba\ further employs a number of state of the art sub-grid physical processes to form realistic galaxies.  
Photoionisation heating and radiative cooling are implemented using the  {\sc grackle}-3.1 library\footnote{\tt https://grackle.readthedocs.io/} \citep{smith-2017} assuming ionisation but not thermal equilibrium, \romeel{with collisional ionisation rates for H taken from \citet{abel-1997} and recombination rates from \citet{hui-gnedin-1997}.}\ {\sc Grackle} accounts for a \citet{haardt-madau-2012} ionising background modified to account for self-shielding based on the \citet{rahmati-2013} prescription (A. Emerick, priv. comm.).  The strength of the ionising background has a very weak impact on the gas dynamics during the simulation, hence it is possible to meaningfully vary this assumption in post-processing without introducing significant errors~\citep{katz-1996}.
The production of 11 different elements (H, He, C, N, O, Ne, Mg, Si, S, Ca, Fe) are tracked, from Type II and Ia supernovae and stellar evolution.  \simba\ tracks dust growth and destruction on the fly, for each individual element \citep[a detailed investigation of the dust model can be found in][]{li-2019}. Star formation is based on a Kennicutt-Schmidt Law~\citep{kennicutt-1998} scaled by the H$_2$ fraction, which is calculated for each particle using its local column density and metallicity following \citet{krumholz-2011}. Galactic outflows are implemented as kinetic decoupled two-phase winds, as in \mufasa~\citep{dave-2016}, with an updated mass-loading factor based on particle tracking results from the Feedback in Realistic Environments (FIRE) zoom simulations~\citep{angles-alcazar-2017b}.  For more details on these implementations, see \citet{dave-2019}.

\subsection{Black hole accretion and feedback}

The energy release from black holes, i.e.
AGN feedback, has a significant impact on the properties of the galaxy and surrounding matter~\citep{fabian-2012}.  \simba\ is notably unique in its way of modelling black hole processes. Owing to the importance of \simba's black hole growth and feedback model for this study, we describe it more detail here; further details are available in \citet{dave-2019}.

\simba\ employs a unique two-mode black hole accretion model.  Cold gas ($T < 10^5$K) is accreted via a ``torque-limited" sub-grid model that captures how angular momentum loss via dynamical instabilities limits gas inflows into the region near the black hole \citep{hopkins-2011,angles-alcazar-2017a}. Meanwhile, hot gas is accreted following the \citet{bondi-1952} formula. The torque-limited mode is appropriate for when black holes are growing in a cold rotationally-supported disk, while Bondi mode is more appropriate for hot gas since it models gravitational capture from a dispersion-dominated medium.  \simba's accretion model thus represents a step up in realism as opposed to simply using Bondi accretion for all forms of gas, as most other current simulations do.  This unique black hole accretion model underpins the implementation of AGN feedback in \simba.

As material accretes into the central region, \simba\ assumes that 10\% of it falls onto the black hole; this accretion efficiency is calibrated to match the amplitude of the black hole mass--galaxy stellar mass relation \citep{angles-alcazar-2013, angles-alcazar-2017a} for massive galaxies from \citet{kormendy-2013}. Accreted gas elements are subtracted a fraction of their mass and immediately ejected as AGN feedback such that the desired momentum flux in the wind ($20L/c$, where $L=0.1\dot{M}c^2$) is achieved. This ejection is purely kinetic, and purely bipolar -- i.e. it is ejected in the $\pm${\bf L} direction where {\bf L} is the angular momentum vector of the inner disk (i.e. the 256 nearest neighbours to the black hole).  \romeel{The physical motivation and detailed implementation for \simba's kinetic AGN feedback are described more extensively in \citet{dave-2019} and \citet{thomas-2019}, but we recap the key points below.}

There are two modes for this type of feedback: radiative mode feedback, and jet mode feedback. The radiative mode in \simba\ happens when there is a high relative accretion rate around a black hole, above a few percent of the Eddington rate. In this mode, the ejected material is kicked with speeds typically around 1000 km/s, scaled to follow observations of ionised gas outflows from \citet{perna-2017a}, and its temperature is not changed in order to represent a multi-phase outflows as observed. At lower Eddington ratios, the jet feedback mode begins to switch on, with full jets achieved below 2\%. The jet mode ejects gas at much higher velocities than the radiative mode, reaching a maximum of $\sim 8000 \kms$. The jet mode also raises the temperature of the ejected particles, based on observations indicating that jets are mostly made of hot plasma \citep{fabian-2012}.  At all times, the amount of matter ejected is mass-loaded from the inner disk in order to have the momentum flux of the outflow be $\approx 20L/c$.  \romeel{This two-mode kinetic feedback broadly follows the physical scenario developed in \citet{best-2012} and \citet{heckman-2014}}.

Besides radiative and jet mode feedback, \simba\ includes also X-ray radiation pressure feedback broadly following \citet{choi-2012}.  This has the effect of pushing outwards on the gas surrounding the accretion disc based on the high-energy photon momentum flux generated in the black hole accretion disk.  It is only activated in low-cold gas content galaxies and when the jet mode is active, because jets tend to be accompanied by strong X-rays and cold dense gas will tend to absorb X-ray energy and radiate it away quickly.

These three forms of AGN feedback -- radiative mode, jet mode, and X-ray -- combine to create a quenched massive galaxy population in good agreement with observations~\citep{dave-2019}, as well as populating them with black holes as observed~\citep{thomas-2019}.  The jet mode is primarily responsible for quenching, although the X-ray feedback has a non-negligible impact. Radiative mode, meanwhile, has a minimal effect on the galaxy population.

\subsection{\simba\ runs}

The \simba\ simulations analyzed in this paper are run in a cubic box with length $50 \hmpc$, with $2 \times 512^3$ elements.  We employ these runs and not the full-size $100\hmpc$ run with $2 \times 1024^3$ from \citet{dave-2019} because we have variants at this box size that enable direct tests of the impact of assumed input physics, particularly AGN feedback. Owing to computational cost, we do not have such variants for the full \simba\ run.  Nonetheless, for all checked properties, the $50\hmpc$ and $100\hmpc$ \simba\ runs agree very well.  \simba\ assumes a cosmology consistent with \citet{planck-2016} results: $\Omega_m = 0.3, \Omega_\Lambda = 0.7, \Omega_b = 0.048, H_0 = 68 \kms {\rm Mpc^{-1}}, \sigma_8 = 0.82$, and $n_s = 0.97$.  The resulting mass resolution is $1.82 \times 10^{7}\msolar$ for gas elements and $9.6 \times 10^{7}\msolar$ for dark matter particles.  

We run several variants of AGN feedback, turning off one input physics quantity at a time, denoted as follows:
\begin{itemize}
    \item ``\simba" denotes a run with all forms of AGN feebdack on.  
    \item ``No-X" denotes a run turning off only X-ray AGN feedback.
    \item ``No-jet" denotes a run turning off both jet and X-ray feedback.
\end{itemize}
We also have a run where all AGN feedback is turned off (``No-AGN"), but it turns out the results are indistinguishable from the No-jet case, hence for simplicity we do not show it here.  Apparently, the radiative  portion of AGN feedback has little impact on the \lya\ forest.  The other three runs allow a direct quantification of the effects of the jet and x-ray AGN feedback modes in \simba.  All these runs are started with identical initial conditions.

We will also compare to the \mufasa\ simulation, the predecessor to \simba\ which does not contain black holes or an explicit AGN feedback model, but rather utilised a heuristic model in which hot halo gas was prevented to cool in order to quench galaxies as observed~\citep{dave-2016,dave-2017b}.  This also employed a $50 \hmpc$ box size with $2 \times 512^3$ elements, with identical initial conditions to the \simba\ runs.

\subsection{Generating spectra}
\label{sec:spectra-and-fgam}

To generate spectra, we employ {\sc Pygad}\footnote{\tt https://bitbucket.org/broett/pygad} \citep{roettgers-2020}.  {\sc Pygad} is a full-featured toolkit for analysing particle-based simulations, including creating mock spectra in any desired ion.  To generate \HI\ spectra, {\sc Pygad} computes the neutral hydrogen fraction for each gas element based on an input (spatially-uniform) UVB \romeel{via a {\sc Cloudy} lookup table (including both collisional and photo-ionisation) interpolated to the redshift of the snapshot}, puts that gas element into velocity space, smooths its neutral component into velocity bins along a chosen line of sight \romeelnew{using the same cubic spline kernel employed in \simba}, and computes the resulting optical depth in each bin.  It further computes the optical depth-weighted density and temperature of \HI\ absorbing gas.  For these spectra, we use a velocity-space pixel size of $6 \kms$.  \romeel{The procedure closely follows what is done in the {\sc Specexbin} code presented in \citet{oppenheimer-2006}, and {\sc Pygad} has been checked to give essentially identical results.}
% Since we do not do line fitting and only consider the \lya\ mean flux decrement $D_A$ in this work, it is not necessary to smooth the spectrum with an instrumental line spread function or to add noise, since these would not change $D_A$.  Note that we do not apply any continuum-fitting to our simulated spectra, which could in principle affect $D_A$. However, at the low redshifts we consider, continuum fitting in observations is usually very accurate owing to the sparse nature of \lya\ forest absorbers, so we assume this has been done accurately in the data that we will compare to.

We generate 1000 spectra for each simulation snapshot through the entire box accounting for periodic boundary conditions.  
\romeel{We apply a line spread function (LSF) for the Cosmic Origins Spectrograph G-130M grating interpolated to each redshift, since COS data will provide our main comparison sample.  We include Gaussian noise with a signal-to-noise ratio of }\jacob{SNR$=12$ per pixel, equivalent to the SNR=$20$ per resolution element that is typical of the COS Guaranteed Time Observations (GTO) team data presented in \citet{danforth-2016}. Finally, we apply a continuum fitting procedure broadly following that described in \citet{danforth-2016}: We obtain the median flux value all pixels in a given spectrum, remove those that are $>2\sigma$ below that median (where $\sigma$ is the inverse of the SNR), and then re-fit the remaining pixels, and iterate until convergence in the median at $<10^{-4}$ relative to the previous iteration.}
%\jacob{(Should we mention that this continuum fitting routine is available in pygad? Also should we mention the particular COS\_G130M instrument?)}

From these spectra, the mean flux decrement $\DA$ was calculated using 
\begin{equation}
    \label{eq:DA}
    \DA = \Big\langle \sum\limits_{i}^{} \left[1-\exp{(-\tau_{i})}\right] \Big\rangle,
\end{equation}
where $\tau_{i}$ is the optical depth in velocity bin $i$ of a given spectrum, and the average is taken over all 1000 generated spectra.

Since \lya\ forest gas is optically-thin, the optical depth of any pixel to good approximation scales as $\tau \propto \frac{1}{\gHI}$. This means any adjustment to $\gHI$ can be related to an adjustment in $\tau$.  This then gives us a way to constrain $\gHI$ using the observed value of $D_A$.  To do this, we multiply $\gHI$ (e.g. from \citealt{haardt-madau-2012}) by a value we denote $\fgam$, which corresponds to multiplying each value of $\tau_i$ by $1/\fgam$; in practice, we do the latter, since optically thick absorption is extremely rare and does not contribute significantly to $\DA$.  The value of $\fgam$ was then adjusted iteratively until the value of $\DA$ computed via equation~\ref{eq:DA} matched the observational determination from \jacob{the combined data of}\ \citet{danforth-2016} \jacob{and \citet{kirkman-2007}}\ (see equation~\ref{eq:DA_obs}) to within 0.0001, at each snapshot redshift.  $\fgam$ can be regarded as the "photon underproduction factor" -- i.e., the amount by which $\gHI$ must be increased in the simulations (assuming a given photo-ionising background) in order to match the observed $\DA$. \jacob{This will be a useful metric for us to quantify the PUC in this work.}

\subsection{Sample mock spectra}

\begin{figure*}
\centering
{
    \includegraphics[width=0.95\hsize]{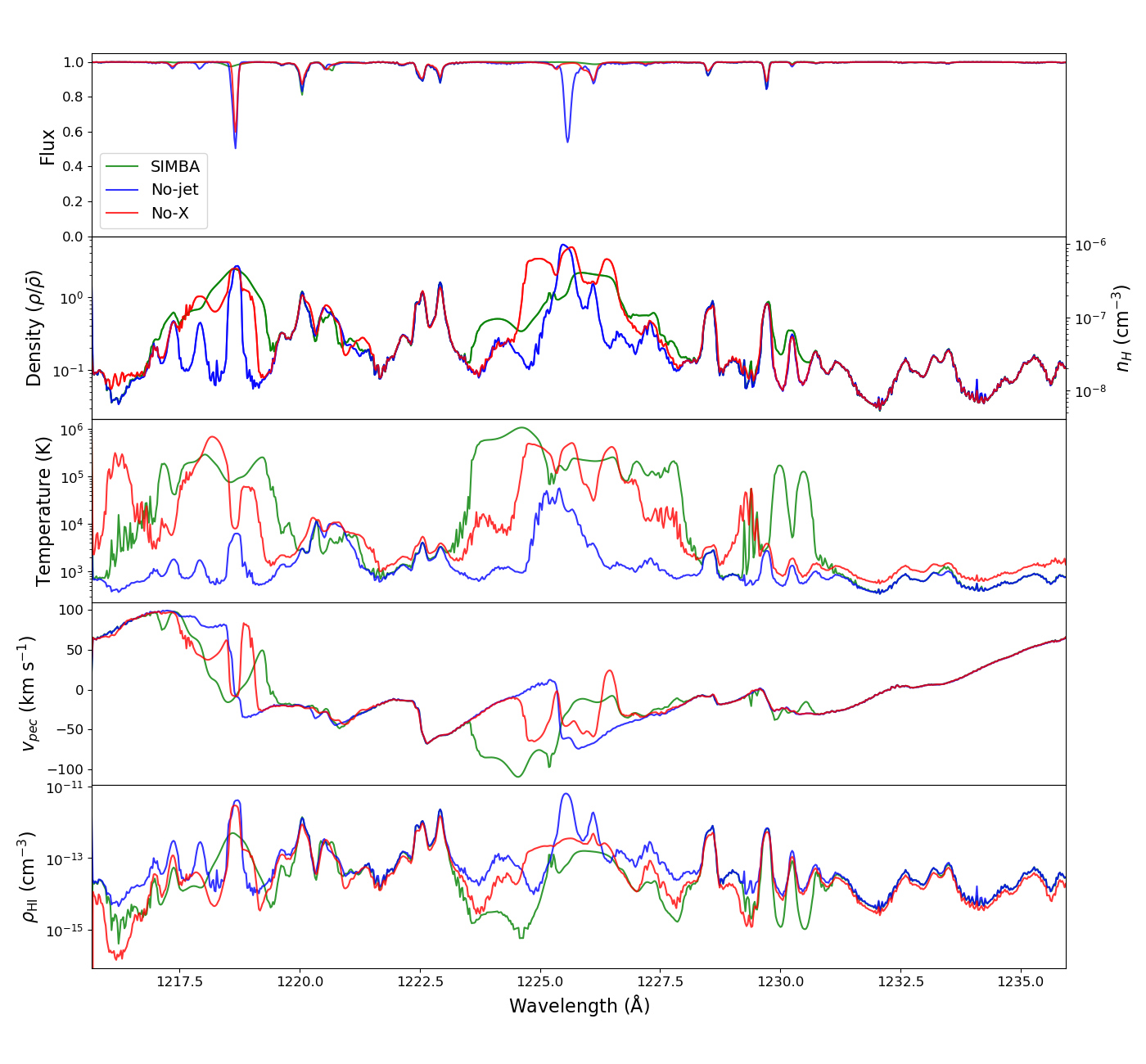}
  }\par\medskip
\caption{An example of three spectra generated using {\sc Pygad}, down the same line of sight at $z=0$: one from the \simba\ simulation with jets turned on (green line), one from the No-jet run (blue line), and one from the No-X run (red line). 
%The top panel shows the flux, the middle panel shows the density normalized to the cosmic mean, and the bottom panel shows the temperature. 
\jacob{The 5 panels, starting from the top show: 
%\begin{enumerate}
%1) Flux, following the process of applying a line-spread function, adding noise, and undergoing continuum fitting;
1) Flux, directly calculated from the optical depths;
2) Gas density, normalized to the cosmic mean (baryonic overdensity);
3) Temperature;
4) Peculiar velocity;
5) Density of neutral hydrogen}. All \jacob{5}\ quantities are plotted in wavelength space. It can be seen that high density gas at low temperatures results in absorption.
%\end{enumerate}
 }
\label{fig:spectra-example}
\end{figure*}

Figure~\ref{fig:spectra-example} shows some example $z=0$ mock spectra generated using {\sc Pygad}. These spectra were all generated down the same line of sight, from our three \simba\ variants:  one from the \simba\ simulation with jet feedback enabled (green), one from the No-jet simulation with jet and X-ray feedback turned off (blue), and one from the No-X simulation with jets enabled but with X-ray feedback disabled (red). \jacob{The 5 panels, starting from the top show: 
%\begin{enumerate}
%1) Flux, following the process of applying a line-spread function, adding noise, and undergoing continuum fitting;
1) Flux (here shown directly calculated from optical depths without any post-processing such as noise being added, to better facilitate comparisons between the models); 
2) Gas density, normalized to the cosmic mean (baryonic overdensity);
3) Temperature;
4) Peculiar velocity;
5) Density of neutral hydrogen}.

%\jacob{The top panel shows the flux, the middle panel shows the baryonic overdensity ($\rho/\overline{\rho}$), and the bottom panel shows the temperature in Kelvin; these quantities are all plotted versus wavelength, and the latter two are weighted by the \HI\ optical depth.}

At $z=0$, the \jacob{flux panel}\ shows that the \lya\ forest is quite sparse compared with higher redshifts, but a number of absorption lines are still visible. Not all of these features are strong enough to be detectable with existing instruments, but this gives an impression of what the underlying HI distribution is within the variants of the \simba\ simulation, without any noise or instrumental broadening.

The \jacob{temperature panel}\ shows that the temperatures are much higher in some parts of the simulations with the jets turned on (\simba\ and No-X) than when they are turned off (No-jet). This illustrates how AGN jet feedback provides an extra source of heating that permeates a significant fraction of the IGM. The additional heating means that the fraction of neutral hydrogen in those regions will be dramatically reduced, and hence that there will be much less \lya\ absorption.  The densities are also significantly impacted, as the higher temperatures result in smoothing the density distribution.

The \jacob{panels show}\ that in some regions, the spectra appear to be almost identical for all feedback variants.  These regions are probing portions of the simulation that have not been affected by jets.  The regions that are affected also usually seem to be relatively denser, which owes to the fact that AGN (and hence AGN feedback) are in galaxies that are biased towards the denser regions.  However, the lowest density regions e.g. towards the right of the spectrum are also unaffected, presumably because they are too far away for jet feedback to have reached there.

Comparing the green and red lines that differ by the inclusion of X-ray feedback, we see that this form of feedback has a small but non-negligible impact on IGM gas heating.  Turning on X-ray feedback (green line) tends to create a slightly more widespread temperature increase around the densest regions, which are presumably closest to galaxies.  The stronger absorption feature around 1219\AA\ in particular shows an interesting case where the X-ray feedback actually has a bigger impact on the absorption than the jet feedback.  This is somewhat unexpected, but it shows that X-ray feedback, despite being explicitly confined to dense ISM gas, still provides an energy input that can somewhat impact larger scales.  Nonetheless, it is clear the primary impact on the density and temperature structure, and hence IGM absorption, occurs due to the inclusion of jet feedback.  In subsequent sections we will quantify these trends in our ensemble of spectra, and use this to understand the implications for the PUC.

\section{IGM physical properties}
\label{sec:igm-background}

We begin by examining some global properties of the IGM in \simba, particularly related to the evolution of the diffuse IGM gas that predominantly gives rise to the \lya\ forest.

\subsection{Visualising IGM jet heating}

\begin{figure*}
\centering
{

  \includegraphics[width=0.95\hsize]{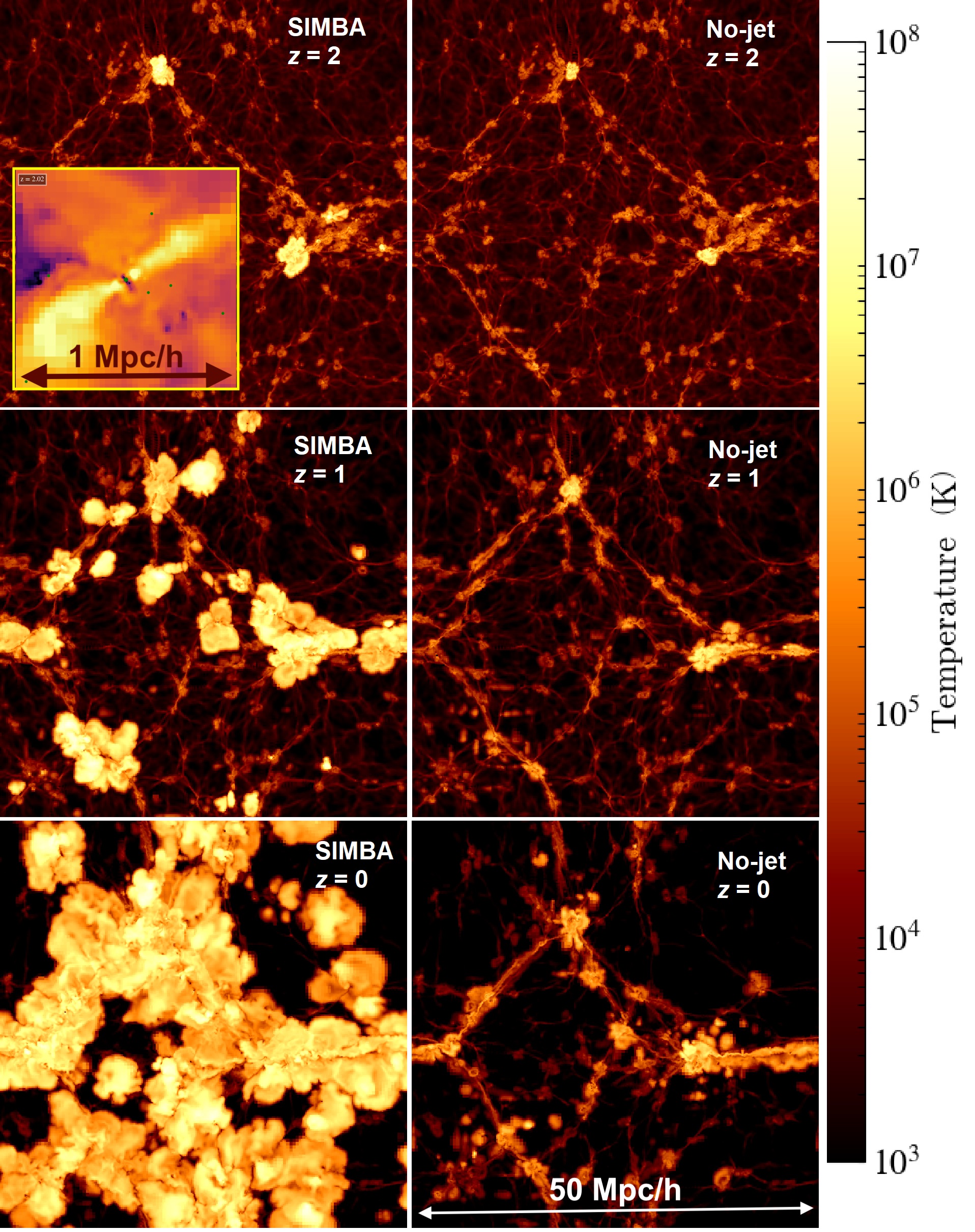}
  }\par\medskip
\caption{$50\hmpc\times 50\hmpc$ temperature slices from \simba\ simulations with AGN jet feedback (left column) and from the No-jet run (right colum). \romeel{Top to bottom rows show $z=2,1,0$, respectively. The inset at $z=2$ shows a $1\hmpc\times 1\hmpc$ zoom on a massive $z\sim 2$ galaxy with a jet, showing sustained bipolar feedback.}\ 
By $z=0$, the jet feedback clearly has a dramatic effect on the temperature of the IGM by, with many Mpc-scale regions heated by jet energy.}
\label{fig:temp-slices}
\end{figure*}

Figure~\ref{fig:temp-slices} shows $50\times 50\hmpc$ temperature maps from our simulations at $z=2,1,0$. The left panels show the full \simba\ run, while the right panels show the No-jet run. The brightest regions represent $T\ga 10^7$K, and the darkest regions down to temperatures approaching a few times $10^3$K that is set by pure photo-ionisation heating.  These images are obtained by computing the mean temperature in each pixel on the $y-z$ plane through the middle of the simulation volume (i.e. at $x=25\hmpc$), using {\sc yt}'s {\tt slice} function. \jacob{There is also an inset on the $z=2$ \simba\ panel (top-left), showing a $1\hmpc \times 1\hmpc$ zoom on a massive $z\sim 2$ galaxy with a jet, showing the bipolar features of the feedback.}

Large-scale filamentary structures are clearly visible in both simulations. These structures stand out as being somewhat hotter than the voids owing to the density--temperature relation in the diffuse photo-ionised IGM~\citep{hui-gnedin-1997}.  Around denser structures, there is additional shock heating caused by gravitational collapse onto filamentary structures, which raises temperatures to $T\ga 10^5$~K. As the simulations evolve to lower redshifts, many of the smaller filamentary structures drain into the larger ones owing to the hierarchical growth of structure, and the IGM is generally cooler owing to its lower physical density and the lower $\gHI$.

Comparing the left and right panels with and without jets, it can be seen that there is only slightly more heating at $z=2$ for simulations with the jets included.  In the jet run, individual bipolar jets are visible around the largest objects, as these generally have the largest black holes and hence low Eddington ratios that transition into jet mode~\citep{thomas-2019}.  The No-jet simulation also has some heating owing to gravitational shock heating in large halos as well as weak feedback.  In general, there are not large differences in the large-scale thermal structure at $z=2$ with the inclusion of jets.

The differences become more drastic at lower redshifts. The No-jet simulation shows heating close to the filamentary structures owing to accretion shocks around large halos, but this heating does not extend very far out. In contrast, the full \simba\ simulation including jets shows heating at $\gg$Mpc scales away from galaxies, which is consistent with the very high velocities at which these wind particles are ejected. For instance, an unimpeded 8000$\kms$ jet will travel $\approx 8$~Mpc in a Gyr. While gravity and interactions with surrounding gas will retard this, it is still plausible that such jets will impact gas out to many Mpc over cosmic time \citep{borrow-2019}.  At $z=0$, many of the locations where the No-jet simulation has cold, diffuse IGM, \simba\ has very hot gas typically in the $T\sim 10^6-10^7$K range.  This clearly demonstrates that jet feedback in \simba\ can have widespread impact in the IGM.

\subsection{Cosmic phase diagram}

\begin{figure}
    \centering
    \includegraphics[width=\linewidth]{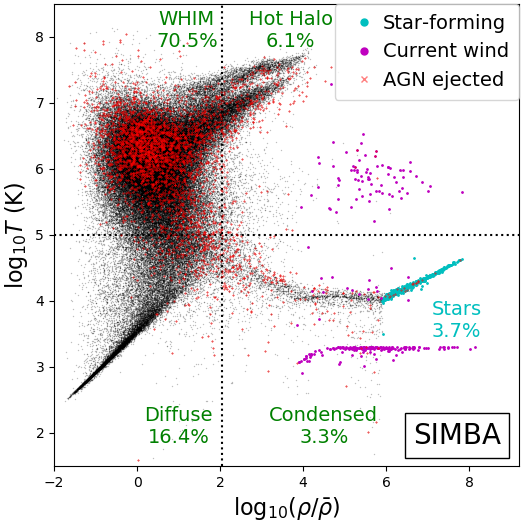}
    %\vskip-0.1in
    \includegraphics[width=\linewidth]{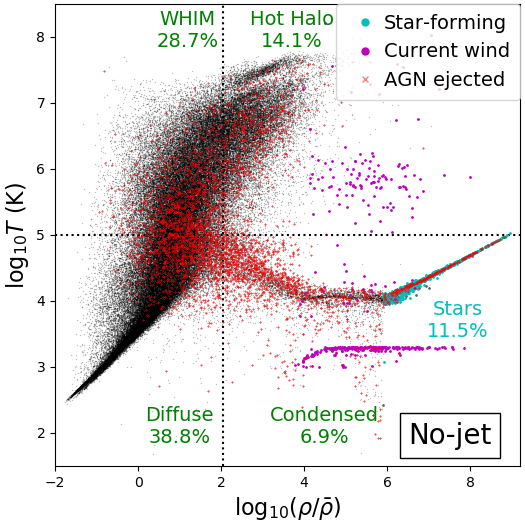} 
    %\vskip-0.1in
    \caption{Phase diagrams at $z=0$ for $50\hmpc$ \simba\ simulations, for the full \simba\ run including jets (top panel) and for the No-jet run (bottom panel). A randomly-selected 0.1\% of gas elements are shown for clarity, as black points. Red points are gas elements that have  at some point been ejected via AGN feedback; this includes from non-jet (radiative mode) AGN feedback. Magenta points are elements which are currently in a decoupled wind, owing to star formation feedback. Cyan points show star-forming gas.  The dotted lines indicate the boundaries between cosmic phases (cf. Figure~\ref{fig:baryon-fractions}):  The vertical division is the approximate density at the virial radius of dark matter halos, while the horizontal division at $T=10^5$K separates cool from warm/hot phases. Percentages of baryons in each phase are indicated. AGN jet feedback results in AGN-ejected particles reaching much further into voids while entraining diffuse gas, thus generating substantially more hot gas well outside of galaxy halos and causing a strong reduction in the amount of cool diffuse IGM gas.}
\label{fig:phase-diagrams}
\end{figure}

An illustrative global diagnostic for understanding IGM evolution is the cosmic phase diagram, i.e. gas temperature versus density of all baryons.  In phase space, gas broadly divides into four regimes~\citep{dave-2010}: {\it Condensed} gas that is cool and dense gas within galaxies and the circum-galactic medium, typically seen neutral and molecular gas; {\it Hot halo} gas that has been shock heated typically to near the halo virial temperature, typically observable via X-ray emission; {\it Diffuse} gas that is mostly photo-ionisation heated in the IGM, which gives rise to the \lya\ forest; and {\it Warm-Hot Intergalactic Medium (WHIM)} gas that has been shock heated to higher temperatures, and which hosts the so-called missing baryons~\citep{dave-2001a}.

Figure~\ref{fig:phase-diagrams} shows the $z=0$ cosmic phase diagram for \simba\ (top panel) and the No-jet (bottom) simulations.  The density has been scaled by the cosmic mean baryonic density. The black points show a randomly selected 0.1\% of the gas (to avoid saturation).  Cyan points show gas that is currently star-forming. Magenta points show gas elements that have recently been ejected in a galactic outflow, and are currently decoupled from hydrodynamics; note that the temperatures of these particles are arbitrary, as they do not currently experience pressure forces.  Finally, the red points show gas elements that have been ejected by either radiative and/or jet AGN feedback at some point in their history.

We divide the phase diagram into four regions, demarcated by the horizontal and vertical dotted lines.  The temperature cut is set at $T=10^5$K, which is a temperature that cannot be obtained without shock heating or feedback, and the traditional definition of the WHIM~\citep{cen-ostriker-1999}.  The density threshold follows \citet{dave-2010} as an estimate of a typical overdensity relative to $\Omega_m$ at the virial radius~\citep[based on][]{kitayama-1996}, given by:
\begin{equation} 
    \label{eq:density-threshold} 
    \delta_{\rm th} = 6 \upi^{2}(1 + 0.4093 (1 / f_{\Omega} - 1)^{0.9052})-1,
\end{equation}
where $f_{\Omega}$ is given by
\begin{equation} 
    \label{eq:fomega}
    f_{\Omega} = \frac{\Omega_{m}(1+z)^{3}}{\Omega_{m}(1+z)^{3}+(1-\Omega_{m}-\Omega_{\Lambda})(1+z)^{2} +\Omega_{\Lambda}}.
\end{equation}
At $z=0$, this results in $\delta_{\rm th}\approx 105$.  We list the mass fraction of baryons in each of these phases on Figure~\ref{fig:phase-diagrams}, along with the baryon fraction in stars that is not included in any of these gas phases but tends to live in dense regions.

The overall phase diagrams in the two cases are generally similar.  The condensed phase consists mostly of photo-ionised gas at $\sim 10^4$~K, along with dense gas forming stars that in \simba\ is forced to lie along a density--temperature relation that explicitly resolves the Jeans mass.  The wind particles are artificially set to $10^3$~K, but as they do not interact hydrodynamically, their temperature has no impact on their dynamics.  The hot halo gas extends up to $T\ga 10^7$~K and generally lies near the virial temperature of its host halo~\citep[e.g.][]{dave-2008}.  The most massive halo in this box is somewhat anomalously large, giving rise to a distinct clump of high-$T$ gas. The diffuse phase shows the tight density--temperature relation characteristic of photo-heated gas expanding with Hubble flow.  Finally, the WHIM phase shows gas that has been shock heated by filamentary accretion as well as feedback processes.

The most notable difference between the \simba\ and No-jet runs is the large decrease in the baryon fraction in the diffuse phase, and a corresponding increase in the baryon fraction contained in the WHIM, when jet feedback is on. The WHIM increase mostly but not entirely comes from the Diffuse phase; the baryon fraction of every other phase is at least halved in the jet simulation compared to the simulation without jets.

The No-jet simulation has baryon phase fractions that are broadly similar to the fiducial model at $z=0$ in \citet{dave-2010}, which had stellar feedback but did not have any AGN feedback.  Hence non-jet AGN feedback has a fairly minimal impact on the cosmic phase diagram.  We have confirmed this for \simba\ by examining the No-AGN simulation, which is not substantially different than No-jet.

%- The phase diagram for the simulation without jets is broadly similar to the vzw, z=0 phase diagram in dave-2010.
Figure~\ref{fig:phase-diagrams} also indicates which gas elements have been ejected by AGN feedback, as red points. In No-jet, we still have radiative AGN feedback up to $\sim 1000\kms$, which distributes some gas into the diffuse and WHIM phase.  However, it does not strongly change the phase of a significant amount of ambient gas; much of it stays at relatively cool temperatures.

In the full \simba\ run with jets, elements touched by AGN feedback can reach well into the diffuse region. In doing so they create a new feature in the cosmic phase diagram at $T\sim 10^6-7$~K near the cosmic mean density, that is not present in the No-jet run.  This region is actually populated mostly by particles that have not been directly kicked by jet feedback, but rather have been entrained (and heated) by jet-ejected gas \citep{borrow-2019}. Also, in this simulation, very few particles that are ejected by AGN feedback end up in the condensed star-forming gas phases, unlike in the No-jet case.  The reason is that the AGN-touched particles are significantly hotter, so do not have a chance to fall back in to bound systems.  This is an important factor for suppressing star formation in massive galaxies having jet feedback, and is a key preventive feedback mechanism that keeps galaxies quenched.

\begin{figure}
\centering
{
  \includegraphics[width=0.95\hsize]{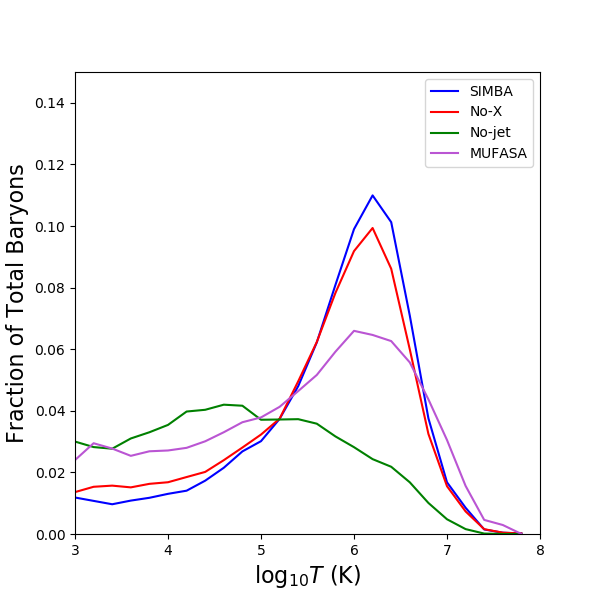}
  }\par\medskip
\caption{Temperature histograms \jacob{at $z=0$}\ of IGM gas ($\rho / \overline{\rho} < \delta_{\rm th}$; i.e. the WHIM and diffuse phases).
Results are shown for $50\hmpc$ simulations with various runs: the main \simba\ simulation (blue line), the No-jet run (green line), the No-X run (red line), and the \mufasa\ simulation (purple line). Including jets (either in \simba\ or No-X) strongly shifts the distribution of IGM gas temperatures, producing a peak at $T\sim 10^{6.2}$K.}
\label{fig:temperature-histograms-of-baryon-fractions}
\end{figure}

Figure~\ref{fig:temperature-histograms-of-baryon-fractions} quantifies the increase in temperature in unbound gas.  It shows histograms of the baryon fraction for low-density phases (i.e. the WHIM and diffuse phases), binned in temperature, for various models.  The most distinct feature is that the \simba\ runs with AGN jet feedback enabled (\simba\ and No-X) have a large peak in their diffuse baryon fractions at $T\sim10^{6.2}$ K.  This shows that jet feedback strongly increases the overall temperature distribution in WHIM gas, compared to the No-jet run (green).  
The \mufasa\ simulations also produce a peak in approximately the same location, but not as sharply; we thus expect that the \mufasa\ simulation will show results intermediate between the No-jet and jet runs. 

When looking at Figure~\ref{fig:phase-diagrams}, remember that the diffuse phase gives rise to \lya\ absorption; the WHIM is too highly ionised for any \HI\ absorption to occur. This means that a decrease in the diffuse fraction will correspond to a decrease in \lya\ absorption.  It is therefore clear that jet feedback will have a significant impact on the amount of \HI\ absorption. This is the primary manner by which AGN jet feedback impacts the \lya\ forest. The extra WHIM gas could potentially generate more high ionisation metal absorption, such as \ion{O}{vi}, \ion{O}{vii}, and \ion{O}{viii}.  Note however that \ion{O}{vi} absorption may not be strongly impacted since \ion{O}{vi} absorption is best at tracing the range $T\approx 10^{5}-10^{5.7}$~K, while Figure~\ref{fig:temperature-histograms-of-baryon-fractions} shows that most of the jet-heated gas is hotter.  Thus \ion{O}{vii} which is strong in $T\approx 10^{5.7}-10^{6.3}$~K gas \citep{nicastro-2018} may be a better tracer~\citep[e.g.][]{chen-2003}. \romeel{ \ion{Ne}{viii} could also provide a useful tracer of $T\sim 10^6$~K WHIM gas in the extreme UV~\citep{burchett-2019}. }

\begin{figure*}
\centering
\includegraphics[width=0.95\hsize]{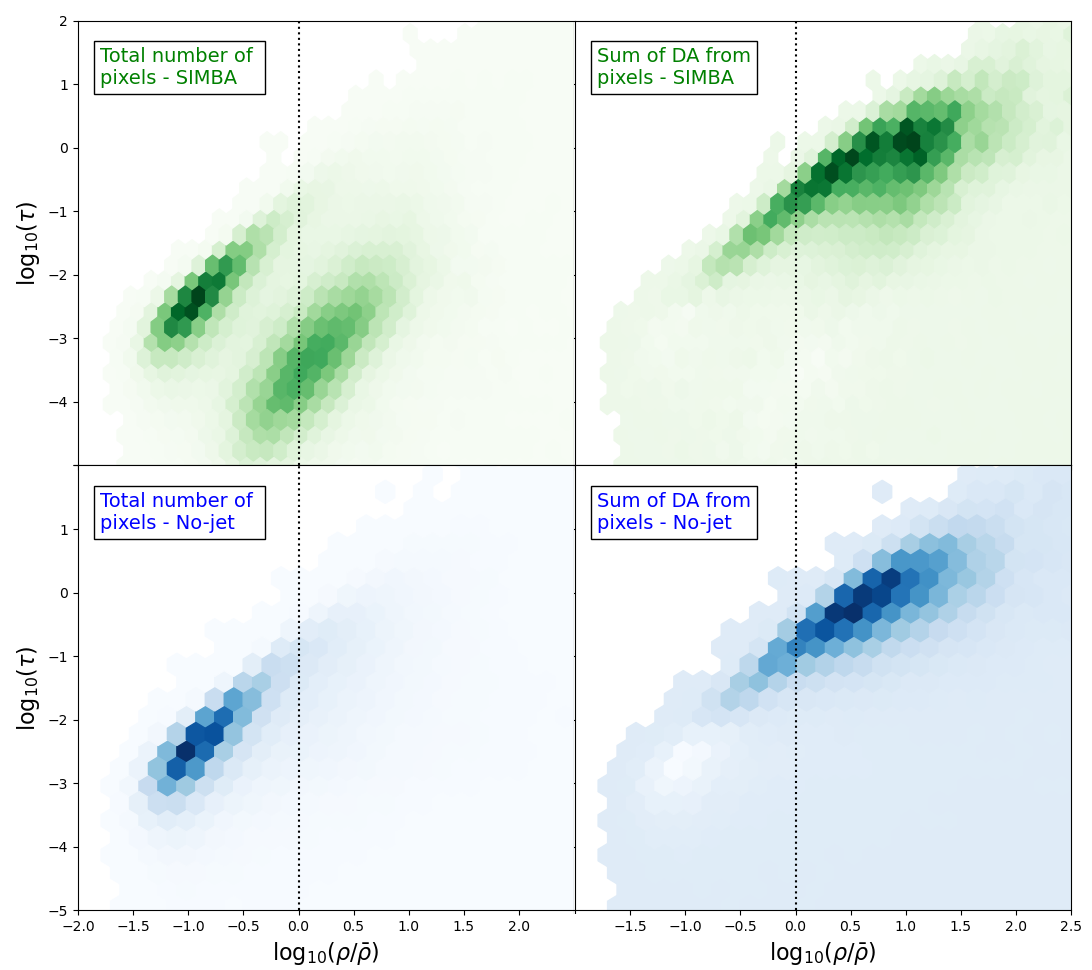}
\vskip-0.15in
\caption{Optical depth $\tau$ vs gas overdensity $\rho/\overline{\rho}$ at $z=0$ for \simba\ (upper panels) and the No-jet Model (lower panels), \jacobnew{showing aggregate properties of}\ all the pixels in a sample of 1000 spectra. \jacobnew{\textit{Left panels}: the total number of pixels with given values of $\rho/\overline{\rho}$ and $\tau$.}\ A large concentration of pixels can be found in the same region for both \simba\ and No-jet, which lies along the
optical depth--overdensity relation set by photoionisation as canonically seen~\citep[e.g.][]{dave-1999}.  But jet feedback in \simba\ has taken a substantial amount of the absorption that occurs around the mean density in the No-jet case, and heated the gas to yield low optical depths. 
\jacobnew{\textit{Right panels:} the sum of $\DA$ from all pixels, at given values of $\rho/\overline{\rho}$ and $\tau$. As expected, in both cases pixels from overdense gas ($\rho/\overline{\rho} > 1$) are the primary contributors to absorption; however, there is a slight change in the distribution of these overdense pixels when jets are included.} }
\vskip -0.1in
\label{fig:tau-vs-rho}
\end{figure*}

\romeelnew{Finally, Figure~\ref{fig:tau-vs-rho} explores how the distribution of pixel flux decrements changes with and without jet feedback.  These show 2-D histograms of optical depth versus overdensity, for \simba\ (including jets) in the top panels, and the No-jet case in the bottom panels.  Left and right panels show the results for unweighted pixels and flux decrement-weighted pixels, respectively.

Looking at the unweighted case (left panels), the vast majority of pixels have low flux decrement, reflecting the sparseness of the \lya\ forest at low-$z$.  The volume-averaged mean overdensity is $\sim 0.3$.  Comparing \simba\ (top) versus No-jet (bottom), we see the emergence of a distinct new cloud of points at moderate overdensities, but low optical depths. These are pixels where jet feedback has heated the gas such that its \lya\ flux has been substantially lowered from where it was in the No-jet case.  This demonstrates that jet feedback is particularly impacting pixels in the moderate overdensity regime ($\rho/\bar\rho\sim 0.5-5$).  This is consistent with the heating of near mean-density gas in the phase diagrams by AGN jets, as seen in Figure~\ref{fig:phase-diagrams}, and will have an impact on the total absorption as discussed in \S\ref{sec:PUC}.

We can see why the impact of jets in this overdensity regime strongly impacts the flux decrements by looking at the right panels of Figure~\ref{fig:tau-vs-rho}, which shows the result of weighting each pixel by its flux decrement.  These panels illustrate where the majority of the flux decrement is coming from in overdensity. Unsurprisingly, the absorption is dominated by flux near $\tau\sim 1$, since at higher $\tau$ values the absorption enters into the logarithmic part of the curve of growth where additional optical depth adds little flux decrement.  In overdensity, $\tau\sim 1$ regions occur at mild overdensities of $\rho/\bar\rho\sim$~few, which from the left plots is exactly the regime that is being impacted by jets.  There is also a clear trend of increasing absorption with overdensity, broadly consistent with previous results~\citep[e.g.][]{dave-2010}.  When including jets, there is a minor but noticeable shift in the peak overdensity, in which the absorption occur over a somewhat broader range in overdensity, shifted to modestly higher values.  There is also an increased scatter at overdensities where jet feedback is impacting the pixel optical depths as shown in the left panels. It is not immediately clear how these trends could be tested observationally, but they may be important in interpreting the underlying cosmic densities traced by low-$z$ \lya\ absorbers.  Overall, this shows that the impact of jets is to heat gas at precisely the range of moderate overdensities that dominate the overall flux decrement, and thus demonstrates why (as we will show later) jets have a significant impact on the flux decrements in the \lya\ forest.
}

%\jacob{In terms of how the entire sample of spectra vary between the different models, Figure~\ref{fig:tau-vs-rho} shows the optical depth plotted against the density for each pixel from a sample of 1000 spectra at $z=0$, for \simba\ and for the No-jet model. The main difference is that a substantial number of pixels in \simba\ have been moved downwards on the y-axis, i.e. they are producing less absorption. This is consistent with the heating of near mean-density gas in the phase diagrams by AGN jets, as seen in Figure~\ref{fig:phase-diagrams}. This has an impact on the total absorption, which will be looked at in \S\ref{sec:PUC}.}

\subsection{Baryonic phase evolution}

Jet feedback clearly has a large impact on the cosmic phase of baryons at $z=0$.  At very high redshifts before jet feedback begins, it should obviously have no impact.  The question is then, when do the \simba\ and No-jet diverge in terms of their baryon fractions in the various phases?

\begin{figure}
\centering
{
  \includegraphics[width=0.95\hsize]{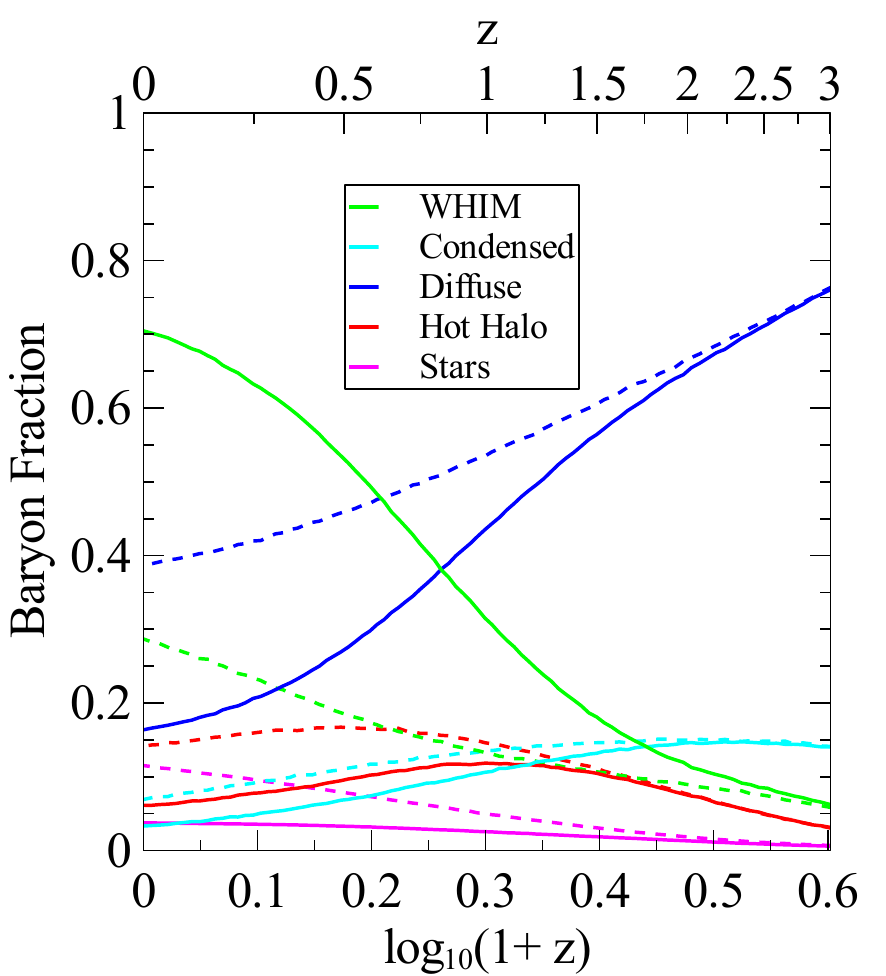}
  }\par\medskip
\caption{Baryon fraction evolution from $z=3\to 0$ in the various phases shown in Figure~\ref{fig:phase-diagrams}, for the $50\hmpc$ \simba\ simulations. The solid lines are from the full \simba\ simulation with jet feedback turned on; the dashed lines are from the No-jet run.  The impact of jets appears at $z\la 2$ and becomes more evident as redshift decreases, with increases in the WHIM phase, and decreases in the diffuse, condensed, hot halo gas, and stellar phases.}
\label{fig:baryon-fractions}
\end{figure}

Figure~\ref{fig:baryon-fractions} shows the evolution from $z=3\to 0$ of the baryon fraction in each phase as defined in Figure~\ref{fig:phase-diagrams}: Green is WHIM, cyan is condensed, blue is diffuse, red is hot halo, and magenta is stars. The dashed lines show the predictions for the No-jet simulation, and the solid lines show the results from the \simba\ simulation with jets.

The simulations both with and without jets have  identical baryon fractions in each phase at $z \sim 3$, since there are essentially no massive black holes with jets yet at these early epochs.
The evolutionary tracks begin to diverge shortly thereafter, with the jet simulation showing more WHIM gas and less in every other phase.  By $z=0$ the jet simulation has almost 2.5$\times$ as many baryons in the WHIM as the simulation without jets, and a corresponding reduction in the diffuse phase. At $z\la 1$ the WHIM phase dominates the baryon fraction in \simba, which never happens in the No-jet case. 

The late onset of these differences is to be expected, as the jet feedback in \simba\ only activates for black holes with masses $\mbh \geq 10^{7.5} \msolar$ with low Eddington ratios, and black holes in \simba\ only reach the required typical sizes at late epochs \citep[see][]{thomas-2019}. The No-jet case broadly reproduces the same evolution of the baryon fractions as the fiducial model used in \citet{dave-2010}, which did not include any AGN feedback.

The \simba\ results with jets show a significantly higher fraction of baryons in the WHIM than previous simulations~\citep{dave-2001a}.  These predicted fractions are also at the high end of current inferences from observations of \ion{O}{vii} absorbers at $z\sim 0.4$, which suggest baryon fractions 20--60\%~\citep{nicastro-2018} in IGM gas with $T=10^{5-7}$~K (see their Table~1).  Our predicted value from the jet simulation is at the top end of this, while from no-jets it is at the bottom end.  We will examine predictions for high-ionisation metal lines from \simba\ in future work, which could be a key discriminant between these types of models with future X-ray missions such as {\it Athena} and {\it Lynx}.

\section{The PUC: Mean Flux Decrement Evolution}\label{sec:PUC}

Armed with an understanding of the physical properties of the IGM, we now examine how AGN feedback impacts \HI\ absorption in the IGM, and thereby investigate the PUC.  To study this, we will use the metric of $\DA$, the mean flux decrement in the \lya\ forest.  This avoids the uncertain and non-unique process associated with line identification and fitting, which can depend fairly sensitively on signal-to-noise, spectral resolution, and other specific aspects that would need to be more closely reproduced in the mock spectra when comparing to observations, and impart greater uncertainties.  For our purposes, $\DA$ provides a robust and well-defined measure that accurately quantifies the PUC.

\begin{figure*}
    \centering
    \includegraphics[width=\linewidth]{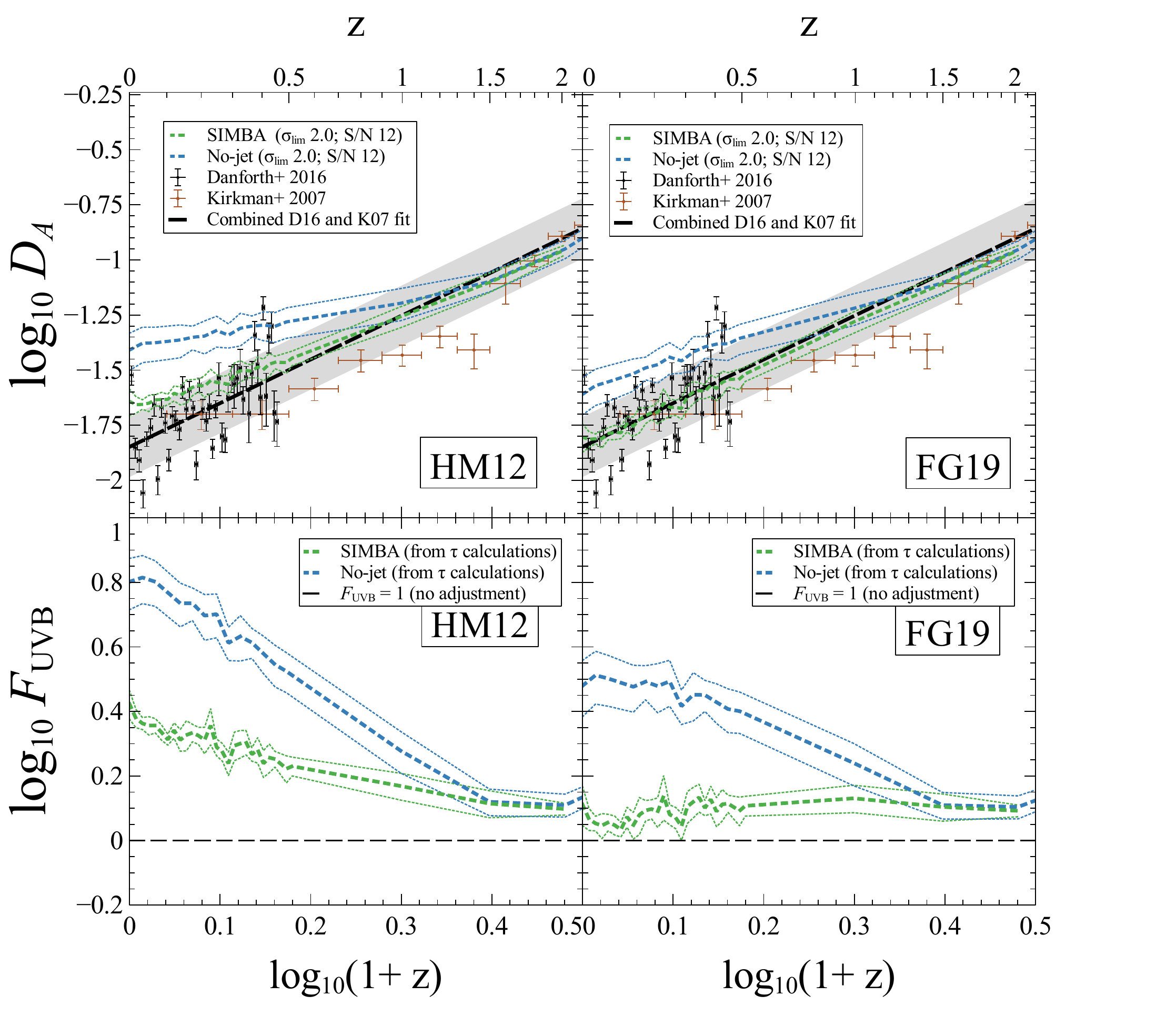}
    \vskip-0.2in
    \caption{{\it Top panels:} $\DA$ versus redshift for simulations versus observations, with the left panels showing the results when simulated spectra are generated using the HM12 background, while right panels show the results when assuming an FG19 background.  The \jacob{dashed}\ green line is from the full \simba\ simulation with jets, while the \jacob{dashed}\ blue line is from the No-jet run; dotted lines indicate estimates of the uncertainty due to cosmic variance.  Black points with error bars are the binned observations from COS data by \citet{danforth-2016} \jacob{, and brown points with error bars show}\ the observational results from \citet{kirkman-2007} from Faint Objects Spectrograph (FOS) data\jacob{, as well as data from KAST and HIRES. The dashed black line shows a combined fit of both observational datasets, with the gray shaded region indicating the uncertainty on the fit.}\
    {\it Bottom panels:} Photon underproduction factor $\fgam$, i.e. the factor by which $\gHI$ must be multiplied in order for simulated predictions of $\DA$ \jacob{(calculated directly from the optical depths, foregoing the continuum fitting process done for $\DA$ in the top panels)}\ to match \jacob{$\DA$ given by the fit to observational data shown in the top panels}. The dashed black line shows the value which indicates no adjustment ($\fgam=1$).
    It can clearly be seen in the top panels that the \simba\ simulations including jets are much closer to matching observed values of $\DA$ than the No-jet runs, regardless of the background used. The FG19 background provides a closer match to observation than the HM12 background. By $z=0$ and using the HM12 background, $\fgam\approx 1.5-2.5$ for \simba, while $\fgam\approx 4-6$ for the No-jet run, showing that jets strongly mitigate the Photon Underproduction Crisis.}
    \vskip-0.1in
\label{fig:DA-and-uvb-factor}
\end{figure*}

Figure~\ref{fig:DA-and-uvb-factor} encapsulates our main results.  Here we show $\DA$ as a function of redshift in the top panels, and the inferred $\fgam$ versus redshift in the bottom panels.  In the left panels, $\DA$ and $\fgam$ have been computed from spectra assuming a \citet{haardt-madau-2012} UVB (henceforth referred to as HM12), while in the right panels a \citet{faucher-2019} UVB (henceforth referred to as FG19) has been assumed. We choose these two backgrounds since the former is the one in which the PUC was originally found, and the latter is a recent state of the art UVB model. The \jacob{dashed}\ green line represents values measured from spectra from the full \simba\ simulation with jets, and the \jacob{dashed}\ blue line represents the No-jet results. Dotted lines indicate uncertainty due to cosmic variance, which was estimated by splitting the spectra into 4 quadrants based on their LOS down the simulation box, and computing the standard deviation on the value of $\DA$ found in each of the 4 quadrants. This cosmic variance uncertainty is typically \jacob{$\sim10\%$}\ for the full \simba\ results, and $\sim20\%$ for the No-jet results.  The effect of cosmic variance appears to be somewhat larger in the No-jet case, which may owe to the fact that without jets, \lya\ absorbing gas is present in highly overdense regions where the variance in absorption is higher, whereas jet feedback removes this. The estimated effect of cosmic variance is in all cases greater than the statistical uncertainty on $\DA$, which is \jacob{$\la 1.0\%$}for all samples.

\romeel{In the top panels, the black data points show the \citet{danforth-2016} measurements from {\it HST}/COS data, while the brown data points show the compilation from \citet[from their Table~5 covering $z\approx 0-3$]{kirkman-2007}.  To get the observed evolution of $\DA$, we fit a single power law to the combined data sets from $z=0-3$:
\begin{equation}
    \log \DA = -1.848 + 1.982 \log{(1+z)}
\end{equation}\label{eq:DA_obs}
We show this as the black dashed line, with a shaded variance computed from the deviations to the individual data points ($\sigma=0.136$~dex).  We will use this line as our baseline observations for comparison to our predicted $\DA$ values.}

The bottom panels of Figure~\ref{fig:DA-and-uvb-factor} show $\fgam$ for the \simba\ simulation variants as a function of redshift. The calculation of $\fgam$ is described in \S\ref{sec:spectra-and-fgam}, and can be regarded as the ``photon underproduction factor'', by which $\gHI$ must be adjusted for simulations to match the observed value of $\DA$. As with the top panels, the bottom-left panel shows the results when using the HM12 background, and the bottom-right panel shows the results when using the FG19 background. \jacob{As $\fgam$ is a rather theoretical quantity, it has been calculated directly from the optical depths, without any continuum fitting being performed.}

Figure~\ref{fig:DA-and-uvb-factor}, top panels, clearly illustrates the PUC.  The No-jet simulation (blue line) shows significantly higher absorption than HST observations, moreso for HM12.  Meanwhile, the absorption in  \simba\ simulation with jets is significantly closer to matching the HST data at low redshifts ($z\lesssim 0.5$), though the HM12 case is still mildly discrepant.  This illustrates our primary result, that including jet feedback and employing a modern determination of the UVB from FG19 essentially solves the PUC in \simba, and allows consistency between source count determined UVB estimates and the estimate obtained from the \lya\ forest.

As we have shown that the jets are a source of additional heating, and heating should reduce the amount of \lya\ absorption, the reduction in $\DA$ with jets is expected. The discrepancy between jet and no-jet results is also expected to be greater going towards lower redshifts, as this is when the jets have had more time to affect the IGM gas in the simulation. At $z\ga 1$, the jet and No-jet simulations do not show strong differences in $\DA$ for either HM12 or FG19, which is expected because there are only minor differences in the diffuse baryon fraction above this redshift (cf. Figure~\ref{fig:baryon-fractions}).  Thus the PUC is only present at $z\la 1$, and increases strongly to lower redshift.

The predicted $\DA$ generally follows a power-law slope in $(1+z)$, but that slope is different depending on whether the HM12 or FG19 UVB is adopted. The slopes when using the FG19 background match the slope of the \citet{danforth-2016} data better than they match the \citet{kirkman-2007} data, while the converse is the case when using the HM12 background. The contrast between the two backgrounds is particularly stark at $z\sim0$.  Fitting a power law with $\DA\propto (1+z)^\alpha$, we obtain slopes of \jacob{$\alpha=[1.5,0.9]$}\ for the jet and No-jet cases respectively for HM12, and \jacob{$\alpha=[1.8,1.3]$}\ for FG19.  This can be compared to the \citet{danforth-2016} slope of $\alpha=2.2\pm 0.2$, showing that the full \simba\ case with FG19 produces a $\DA(1+z)$ slope in very good agreement with observations, and as a result a non-evolving $\fgam$.  At higher redshifts ($z\ga 1$), the predicted values of $\DA$ match fairly well with expectations from either HM12 or FG19, which nicely demonstrates that prior to the impact of AGN jet feedback, the UVB amplitude determined from source count modeling is in good agreement with that inferred from the \lya\ forest.

To more precisely quantify this excess of absorption shown in the top panels, we show $\fgam$ as a function of redshift in the bottom panels of Figure~\ref{fig:DA-and-uvb-factor}.  For the No-jet case and the HM12 UVB, the photon underproduction factor reaches $\sim 6$ at $z=0$, and is already $\sim 3$ at $z=0.5$.  This confirms the PUC found by \citet{kollmeier-2014} in the case with no AGN feedback and HM12.  In fact, even though the No-jet run has some AGN feedback, the underproduction factor is higher compared to the $\times 5$ discrepancy found by \citet{kollmeier-2014}.  This may owe to the lower star formation-driven wind speeds in \simba\ relative to the \citet{dave-2013} simulations used in \citet{kollmeier-2014}, and/or the use of MFM rather than SPH for the hydrodynamics.  In any case, the overall results are very similar, and confirm that the PUC is present in state of the art simulations when no feedback is included that heats the IGM.

With jets on, the green line shows that the PUC is not completely eradicated -- at $z=0$, with HM12, the photon underproduction factor is still 2.5 (lower left panel).  However, this is clearly much closer to unity, which would be the value if the predicted $\DA$ exactly matched the \citet{danforth-2016} measurements.  Given that there are $\sim\times 2$ uncertainties in the source count modeling determinations of $\gHI$~\citep{khaire-2015}, such a discrepancy may not be considered severe.  

Looking at the lower right panel which assumed FG19 instead of HM12, the PUC is essentially gone.  The No-jet case still has a factor of 3 discrepancy in $\fgam$, while the jet simulations reduces this to $\sim1.2$, which is now likely well within current uncertainties. Interestingly, the evolution of $\DA(z)$ predicted in the \simba\ simulation is in very good agreement when assuming FG19, but with HM12 we predict a fairly strongly increasing PUC to lower redshifts. Thus the \simba\ simulations with jet feedback and using the FG19 background are in quite good agreement with the low-redshift \lya\ forest data from \citet{danforth-2016}. 

It is interesting to note that $\fgam$ is actually somewhat larger than the discrepancy in $\DA$ from the top panel.  For instance, at $z=0$ for the No-jet case and HM12 background, the ratio of the predicted $\DA$ (blue line) and the \citet{danforth-2016} value is about a factor of \jacob{3}.  However, when one goes through the exercise of iteratively adjusting the ionising background to match $\DA$, this indicates that a factor of 6 is needed to match the observations.  The reason is that saturated lines provide a sub-dominant but non-negligible contribution to $\DA$.  Saturated lines move into the logarithmic portion of the curve of growth, so their flux decrement no longer scales linearly with optical depth and $\gHI^{-1}$.  Hence it is important to do the exercise of iteratively fitting to the observed $\DA$ as we have done, since the PUC is actually worse that it appears simply by examining the discrepancy in $\DA$.

The impact on $\DA$ at lower redshifts owes not only to the increasing filling factor of hot gas as evident from Figure~\ref{fig:temp-slices}, but also to the fact that the largest contribution to $\DA$ comes from marginally saturated lines ($\nhi\approx 10^{13.5-14}\cdunits$), since below this the column density distribution has a slope shallower than $-2$~\citep{danforth-2016}, and above this the increase in absorbers' column densities no longer contribute linearly to $\DA$.  At $z\sim 2-3$, marginally saturated lines correspond to gas at moderate overdensities of a few, but by $z=0$, these lines arise in diffuse gas of overdensities of $\sim 20-50$~\citep{dave-1999}.  As a result, they move into the regions nearer to galaxies that are most dramatically impacted by the jet heating.  This exacerbates the effect on $\DA(z)$.

As mentioned in \S\ref{sec:igm-background}, the diffuse phase of matter at low densities and temperatures is responsible for \lya\ absorption. The $\times 2.5$ reduction in $\DA$ when jets are turned on as seen in Figure~\ref{fig:DA-and-uvb-factor} is consistent with the $\times 2.5$ reduction in the fraction of baryons in the diffuse \lya-absorbing phase, as seen in Figures~\ref{fig:phase-diagrams} and \ref{fig:baryon-fractions}. In light of this, a straightforward physical interpretation of the impact of AGN feedback on the low-redshift IGM is that it serves to heat a sufficient fraction of diffuse gas into the WHIM phase in order to provide a potential resolution to the PUC.

\section{Flux Probability Distribution Function}\label{sec:fpdf}

\begin{figure}
\centering
\includegraphics[width=0.95\hsize]{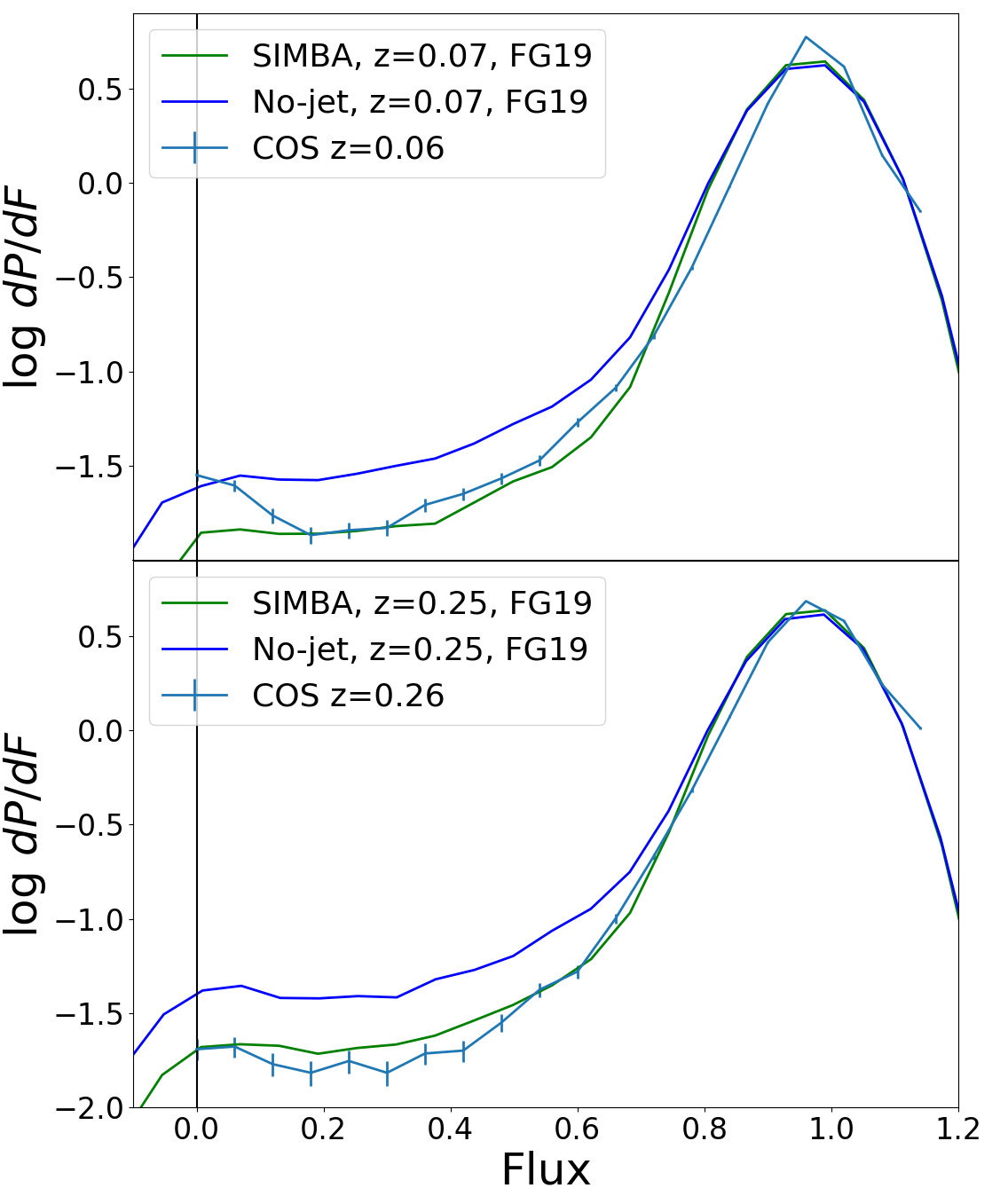}%{fpdf-old.png}
\vskip-0.1in
\caption{\romeelnew{The Flux Probability Distribution Function (FPDF) in \simba\ (green) versus No-jet (blue), compared to observations from the COS GTO team~\citep[cyan;][]{danforth-2016}, compiled as described in the text.  \simba\ provides a very good match to the observed FPDF, much better than the No-jet case, demonstrating that \simba's jet feedback suppresses absorption within flux bins in accord with data.}}
\vskip -0.1in
\label{fig:fpdf}
\end{figure}

\romeelnew{While $D_A$ measures the mean absorption, the distribution of pixel fluxes provides a more detailed test of whether \simba\ yields an accurate description of the low-redshift \lya\ forest~\citep{gaikwad-2017b}. Therefore in this section we examine the flux probability distribution function (FPDF), which is the density histogram of normalized flux values from a set of quasar spectra. The FPDF tests whether the {\it distribution} of pixel flux decrements is in accord with observations, which is a higher order constraint as compared to $D_A$.

To compare to observations, we download the reduced quasar spectra obtained by the COS GTO team \citep{danforth-2016} from the Mikulski Archive for Space Telescopes\footnote{\tt https://archive.stsci.edu/prepds/igm/}. We mask out pixels that have significant foreground contamination, identified as having $>2\sigma$ absorption in the composite foreground spectrum.  For the remaining pixels, we collate the fluxes for all the pixels with rest-frame \lya\ wavelengths of $>1040$\AA\ (to avoid Ly$\beta$ and \ion{O}{vi} absorption), are not within 5000~km~s$^{-1}$ of the quasar systemic redshift, and are $>5000$~km~s$^{-1}$ redwards of $z=0$ in order to avoid Galactic absorption.  For these fluxes, we subtract off any remaining foreground absorption, and normalise the flux using the continuum provided by the COS GTO team.  We then compute the FPDF from this continuum-normalized flux, in two bins of $\Delta\log (1+z)=0.0385$, which correspond to $z=0-0.093$, and $z=0.194-0.305$.  We have checked that the intermediate redshift bin gives very similar answers, and while the sample contains some higher-$z$ data, it is fairly sparse and noisy so does not give useful constraints.  The mean redshifts for the pixels in these bins are $\bar{z}=0.06$ and $\bar{z}=0.26$.

Figure~\ref{fig:fpdf} shows the FPDF (i.e., $dP/dF$ where $P$ is the probability density of a pixel having a flux $F$) in \simba\ (green line) and No-jet (blue), versus the COS GTO data (cyan) from \citet{danforth-2016}. Errors on the COS GTO data are computed by propagating the errors from the individual pixels and continuum level, but do not include cosmic variance, so may be underestimated.  In the simulations, we compute the FPDF at the snapshot whose redshift is closest to the mean redshift of COS GTO pixels in that bin, as indicated in the legend.  For clarity, we only show the FG19 results for the simulations, to focus on how jet feedback impacts the FPDF.

\simba\ provides a very good match to the shape and amplitude of the observed FPDF.  The only significant deviation is seen for near-saturated fluxes ($\la 0.1$) in the lower-$z$ bin, but given that such saturated absorption is rare, it is likely that cosmic variance dominates the uncertainty here. Otherwise, over most of the fluxes, \simba\ is a much better match to the observations than the No-jet case, which yields an FPDF that is $\sim\times 2$ higher for fluxes with significant absorption.

This demonstrates that \simba's jet feedback not only suppresses the mean absorption in accord with observations, but it does so in a way that further yields a pixel flux distribution in good agreement with observations.  This confirms that \simba's jet feedback, together with the FG19 background, provides a much better representation of the low-$z$ \lya\ forest as compared to a simulation that does not include such widespread AGN feedback.
}

\section{AGN feedback variants}\label{sec:variants}

\begin{figure}
\centering
\includegraphics[width=0.95\hsize]{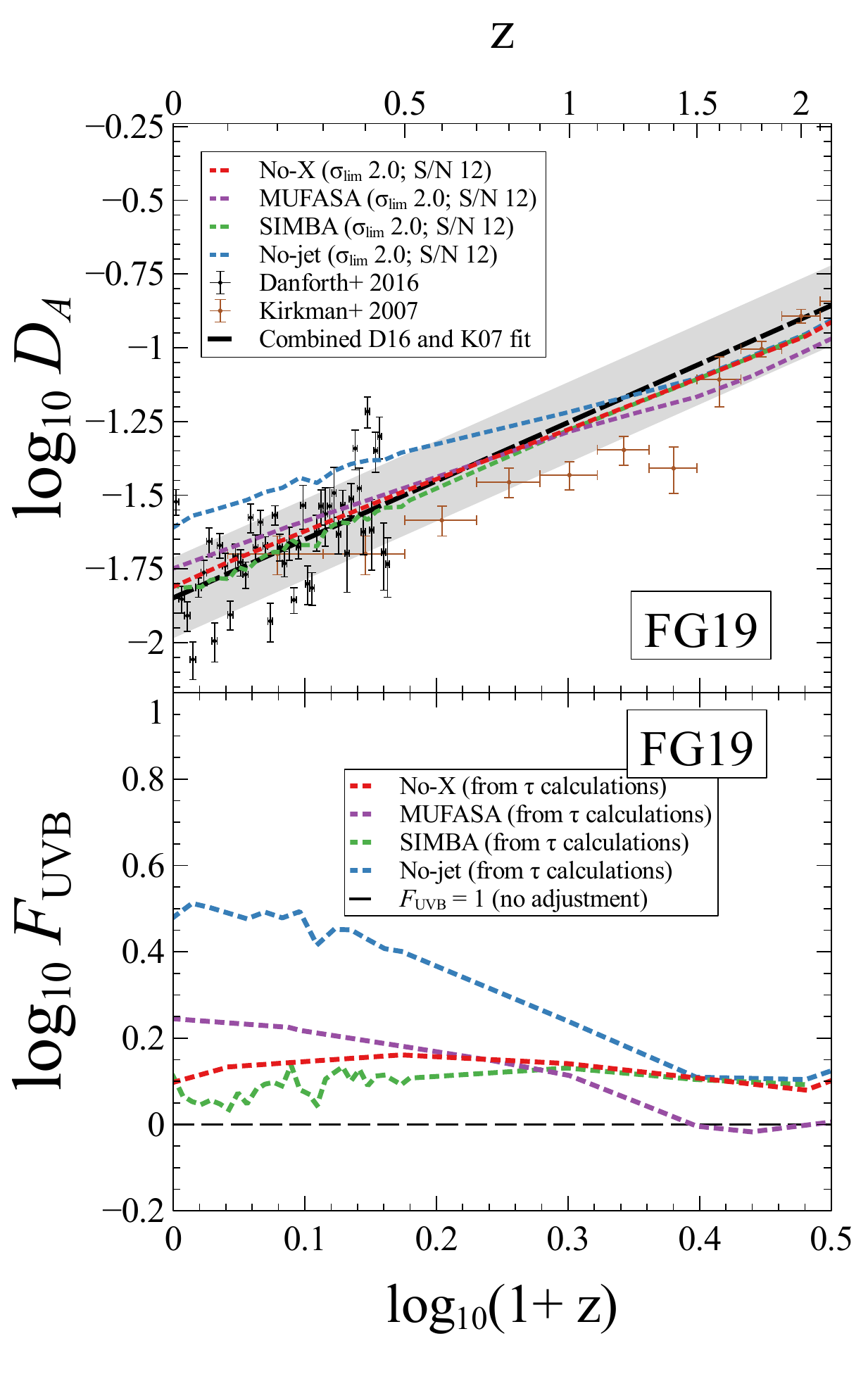}
\vskip-0.2in
\caption{$\DA$ (top panel) and $\fgam$ (bottom) as a function of redshift for various \simba\ simulations using the FG19 background, similar to Figure~\ref{fig:DA-and-uvb-factor}.  The red line shows results from the No-X simulation with jets but without X-ray feedback, and the purple line shows results from the \mufasa\ simulation. Green and blue lines are reproduced from Figure~\ref{fig:DA-and-uvb-factor} showing the \simba\ and No-jet runs for comparison.
Observations are also reproduced from Figure~\ref{fig:DA-and-uvb-factor}, as indicated. The No-X simulation is quite similar to the \simba\ run with X-ray feedback, showing that X-ray feedback has negligible impact on the diffuse IGM, and thus the impact comes from the jets.  \mufasa\ matches observational data better than the No-jet case, but not as well as with jets, indicating that these observations could potentially discriminate between otherwise successful AGN feedback models.}
\vskip -0.1in
\label{fig:DA-mufasa-xray}
\end{figure}

In the previous section, we focused on comparing the full \simba\ simulation with the No-jet run, because these provide the greatest differences illustrating the impact of AGN feedback.  In this section, we further consider two additional model variants, to gain insights into how well these PUC measurements might be able to discriminate between AGN feedback models.  In the No-X case, we have left jets on but turned off X-ray AGN feedback; if jets are the dominant mechanism impacting the IGM, we expect this model to be similar to the full \simba\ run, as opposed to the No-jet run which turns off both jet and X-ray feedback.  We will also consider \mufasa, which used a completely different method for quenching galaxies in which hot gas in halos above an (evolving) mass threshold was prevented to cool~\citep{dave-2016}.  The No-X model produces mostly quenched galaxies but with insufficiently low specific star formation rates compared to observations~\citep{dave-2019}, while \mufasa\ produces a quenched population in very good agreement with observations~\citep{dave-2017b}, in some ways even better than \simba, but it uses a less physical approach that does not directly model black holes.  Here we examine $\DA(z)$ and $\fgam(z)$ in these two variants.

Figure~\ref{fig:DA-mufasa-xray} shows $\DA$ (top panel) and $\fgam$ (bottom) as a function of $\log\ (1+z)$, as in Figure~\ref{fig:DA-and-uvb-factor}.  Here we focus on just the FG19 background, as this one is overall more successful for \simba.   We show the results from No-X and \mufasa\ as the red and purple lines, respectively.  For comparison, we continue to show the \simba\ and No-jet lines in green and blue, respectively.  The observations are also shown as presented in Figure~\ref{fig:DA-and-uvb-factor}. For clarity, the uncertainties due to cosmic variance are omitted from the graph, but are typically \jacob{$\sim10\%$}\ for the No-X case (similar to the full \simba\ run) and up to \jacob{$\sim30\%$}\ for the \mufasa\ results (somewhat higher than the No-jet case). It is not immediately evident why \mufasa\ would exhibit such large cosmic variance, nonetheless this is still relatively small compared to the values needed to solve the PUC. 

Figure~\ref{fig:DA-mufasa-xray} demonstrates that X-ray feedback has a negligible impact on \lya\ absorption in the IGM; the values in red and green are nearly overlapping at all redshifts, though the $\DA$ values are very slightly higher than \simba\ without the additional feedback from X-rays.  This is expected, as the X-ray feedback primarily acts within the inner disk of galaxies close to the AGN, and thus is not expected to directly impact IGM gas. This conclusively demonstrates that it is in particular the AGN jet feedback that is responsible for lowering $\fgam$ in \simba.

For \mufasa, it is interesting to note that there is still a substantial reduction in $\fgam$, moving $\DA$ closer to the observed values, though not as strongly as in \simba. This was anticipated from Figure~\ref{fig:temperature-histograms-of-baryon-fractions}, which showed that \mufasa\ generates a substantial shift in the IGM temperature distribution from that expected with no or weak AGN feedback.  This is somewhat surprising because the direct impact of the feedback is confined to halo gas (by adding heat to offset cooling), yet it appears to have a wider impact on IGM gas.  
Nonetheless, by $z=0$, the photon underproduction factor is still $\approx 2$, so significantly higher than in \simba, though well lower than in the No-jet case.  We do not show the HM12 results here, but the corresponding factor for \mufasa\ in this case is $\approx 4$.  Hence one might envision, with improved measurements of $\gHI$ in the local universe such as from flourescence~\citep{fumagalli-2017}, it may be possible to discriminate between variants of AGN feedback based on their impact on the diffuse IGM.

\jacob{
\section{Modelling Uncertainties}\label{sec:modelling-uncertainties}
%Section about the effect of continuum fitting, LSF, choice of $\siglim$, SNR, boxsize, mass element resolution, etc.
As can be seen in the description of spectra generation in Section~\ref{sec:spectra-and-fgam}, there are many factors which may in principle have an influence on the calculation of $\DA$. These include: the continuum fitting process, and the choice of the free parameter involved; the application of a line-spread function (LSF) to the flux to imitate the effect of light passing through an instrument; the signal-to-noise ratio (SNR) of the noise added to the spectra; the boxsize of the simulation; and the mass element resolution of the simulation, among others. 

%\textit{\textbf{``We show in this section that while these factors have some effect on the value of $\DA$, the effects are mostly minor and do not change the overall conclusions that we have drawn in Section \ref{sec:PUC}.'' (?)}}

Figure~\ref{fig:continuum-sigma-lim-effect} shows $\DA$ for variations in different steps of the spectra generation process. Each of the 4 lines labelled `\simba' is from the primary \simba\ simulation including jets. The \simba\ line labelled `$\tau$ calculations' shows the value of $\DA$ when calculated directly from the optical depths, with no effect from the LSF, added noise, or continuum fitting process. The other 3 \simba\ lines all involve the calculation of $\DA$ using flux that has undergone the application of the COS LSF, and a continuum fitting process; they vary in either the SNR, or in the value of $\siglim$ used in continuum fitting. Specifically, pixels which are $\siglim / {\rm SNR}$ below the median pixel flux are removed during each iteration of the continuum fitting process. The application of the COS instrument LSF was found not to make any discernible difference in the overall $\DA$ of the sample, and thus we only show values calculated with the LSF applied in Figure~\ref{fig:continuum-sigma-lim-effect}.

It can be seen from Figure~\ref{fig:continuum-sigma-lim-effect} that using a value of $\siglim=2.0$ produces a significantly lower value of $\DA$ compared to using $\siglim=1.5$, regardless of the SNR used. A higher SNR also lowers $\DA$, but not as drastically as the effect of $\siglim$. This trend with $\siglim$ is to be expected, as using a larger value of $\siglim$ will result in a lower continuum level. The proper value to set $\siglim$ can be difficult to determine, hence why we provide this comparison.  One way to set this is to see which value results in the closest match to the true value (i.e. `tau calculations').  At $z\la 0.5$, we find that $\siglim = 2.0$ matches reasonably well.  This is also a typical value taken in observations.  However, \citet{danforth-2016} actually used $\siglim = 1.5$, which would tend to raise $\DA$ above the true value.  Note that they were fitting to an intrinsically non-uniform continuum and generally over longer stretches than our spectra using a higher order fitter, hence their choice of $\siglim$ is not directly comparable to ours, but it is illustrative of the changes that can occur owing to this choice.  In our procedure with \simba, such a choice clearly over-fits the continuum at $z=0$.

While we have shown the full \simba\ run here, we have checked that the same trends generally hold in the No-jet simulation.  This means had we used $\siglim = 1.5$, the PUC effect noted in the No-jet simulation would be even more severe than we presented.  It would also mean that even including \simba's jet feedback would be unable to fully solve the PUC.  

%Based on the largest variation from the fiducial values of $\siglim=1.5$ and SNR$=12$, we estimate the modelling uncertainties at $\sim 0.2$ dex.
%\textbf{This uncertainty is not sufficient to explain the discrepancy between the No-jet results and observation when using an HM12 background, but is just exactly enough to bring them on the outer edges of the fit uncertainty when using a FG19 background.}

% ``A lower value of sigma will yield a higher continuum.  Ideally, the value should be chosen so that the points below 1 are true continuum values, just slightly lower owing to noise.  I always thought the canonical value was 2.0, actually.  This is because for e.g. ~1000 pixels, one expects maybe ~25 pixels below 2 sigma low.  This is a small enough number that it should not bias the fit much if one throws those out and re-fits.  But for 1 sigma, it is like 160 pixels, so that is a lot.  1.5 sigma would be about 70 pixels out of a 1000 below the 1.5sigma threshold.  This is why it’s tricky to set this number, it’s not entirely obvious what it should be.  In this case, however, because we are comparing to observations that use sigma=1.5, we should probably just stick with that. ''

%\textit{\textbf{From this, we can estimate a modeling uncertainty that spans the lowest to the highest of these runs; at $z=0$, this reaches approximately $\pm 0.1$ dex. } }

%\textbf{$\pm 0.225 dex$ ? since we're going from the continuum fit line in the }

\begin{figure*}
    \centering
    \includegraphics[width=\linewidth]{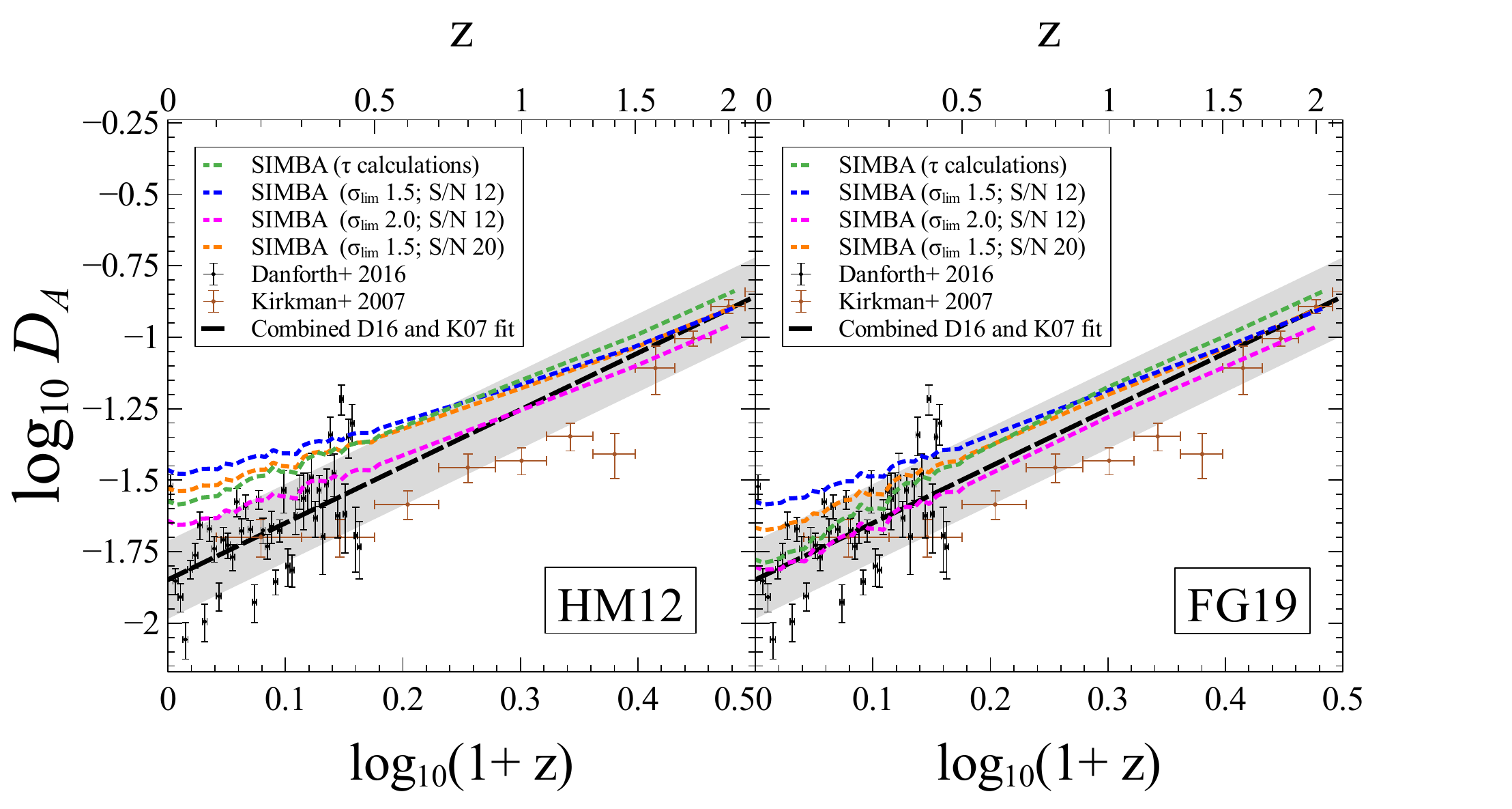}
    \vskip-0.2in
    \caption{\jacob{Graph showing the changes in $\DA$ from spectra in \simba\ resulting from changes in parameters used in the continuum fitting process, assuming the HM12 background (left) and FG19 background (right). The dashed green line shows the \simba\ results when calculated directly from the optical depths, with no continuum fitting process being done. The dashed blue, magenta, and orange lines show the \simba\ results when a continuum fitting process is performed on the spectra, using a SNR of 12 and a value of $\siglim=1.5$, SNR=12 and $\siglim=2.0$, and SNR=20 and $\siglim=1.5$, respectively.  The magenta line is our default choice in presenting our earlier results (e.g. Figure~\ref{fig:DA-and-uvb-factor}).  The observational results from \citet{danforth-2016} and \citet{kirkman-2007}, along with a combined fit of their data are shown identically to previous graphs of $\DA$.}}
    \vskip-0.1in
\label{fig:continuum-sigma-lim-effect}
\end{figure*}

Figure~\ref{fig:m25-comparisons} shows $\DA$ for the primary \simba\ simulation used in this report (labeled `\simba', which as noted previously has a boxsize of $50 \hmpc$ and $512^3$ mass elements), compared with two variants of the \simba\ simulation. One of these variants uses a boxsize of $25 \hmpc$ with $512^3$ mass elements (labeled `m25n512'), and the other uses a boxsize of $25 \hmpc$ with $256^3$ mass elements (labeled `m25n256'). Aside from the boxsize and number of mass elements, the simulation physics for all 3 are identical.

It can be seen in Figure~\ref{fig:m25-comparisons} that both the m25n512 and m25n256 simulations have slightly more absorption than the main \simba\ results; however, both of the simulations using a boxsize of $25 \hmpc$ have almost identical absorption, with the slight exception of right around $z=0$. The variation in parameters between the 3 simulations sheds light on the effects of boxsize and mass resolution on our results for $\DA$. There appears to be a minor increase in absorption at close to $z=0$ when the resolution is increased, but we note that this is insignificant in magnitude compared with the differences between the \simba\ and No-jet simulations seen in Figure~\ref{fig:DA-and-uvb-factor}. 

The fact that both m25n512 and m256n256 are more similar to each other than to regular \simba\ suggests that boxsize has a greater effect on $\DA$ than resolution does, at least among the values under consideration.  The \simba\ simulation with the larger boxsize of $50 \hmpc$ but same resolution appears to have somewhat less absorption than the $25 \hmpc$ simulations at $z<1$. This occurs because the $25 \hmpc$ simulations has fewer large galaxies owing to its non-representative volume, and therefore will have fewer large black holes producing jet feedback. This means that less of the IGM will have been heated, and these simulations will therefore have somewhat higher absorption. %Ideally we would have performed our analysis on the \simba\ simulation mentioned in \citet{dave-2019} with a boxsize of $100 \hmpc$ and $1024^3$ mass elements, but the computational resources for such an analysis were not available for this project.

In short, choices in continuum fitting can have an impact of up to $\sim 0.2$~dex on the predicted values of $\DA$ at $z=0$.  While this is substantial, it is roughly independent of the feedback model, so it does not lower the differences between the Jet and No-Jet. It also cannot fully explain the PUC, without rather extreme choices. If for instance we were to match the \citet{danforth-2016} choice of $\siglim=1.5$, we actually produce an even larger PUC that is not fully mitigated even with the inclusion of jets.  While using a very large value of $\siglim$ could mitigate the PUC, it is a non-standard choice that would give rise to significant a mismatch between the fitted continuum and the true continuum.  In addition, we have checked that our results are unlikely to be significantly influenced by the numerical effects of simulation resolution.  However, they are significantly impacted by box size when jets are included, because jets preferentially occur in massive red and dead galaxies that are underrepresented in small volumes.  %If anything, we expect that the discrepancy between \simba\ and the No-jet model may actually be slightly larger in the $100 \hmpc$ \simba\ simulation than is seen in our results from the $50 \hmpc$ simulation. 
We conclude that our main results in Figure~\ref{fig:DA-and-uvb-factor} are not driven by numerical or analysis artifacts.

\begin{figure*}
    \centering
    \includegraphics[width=\linewidth]{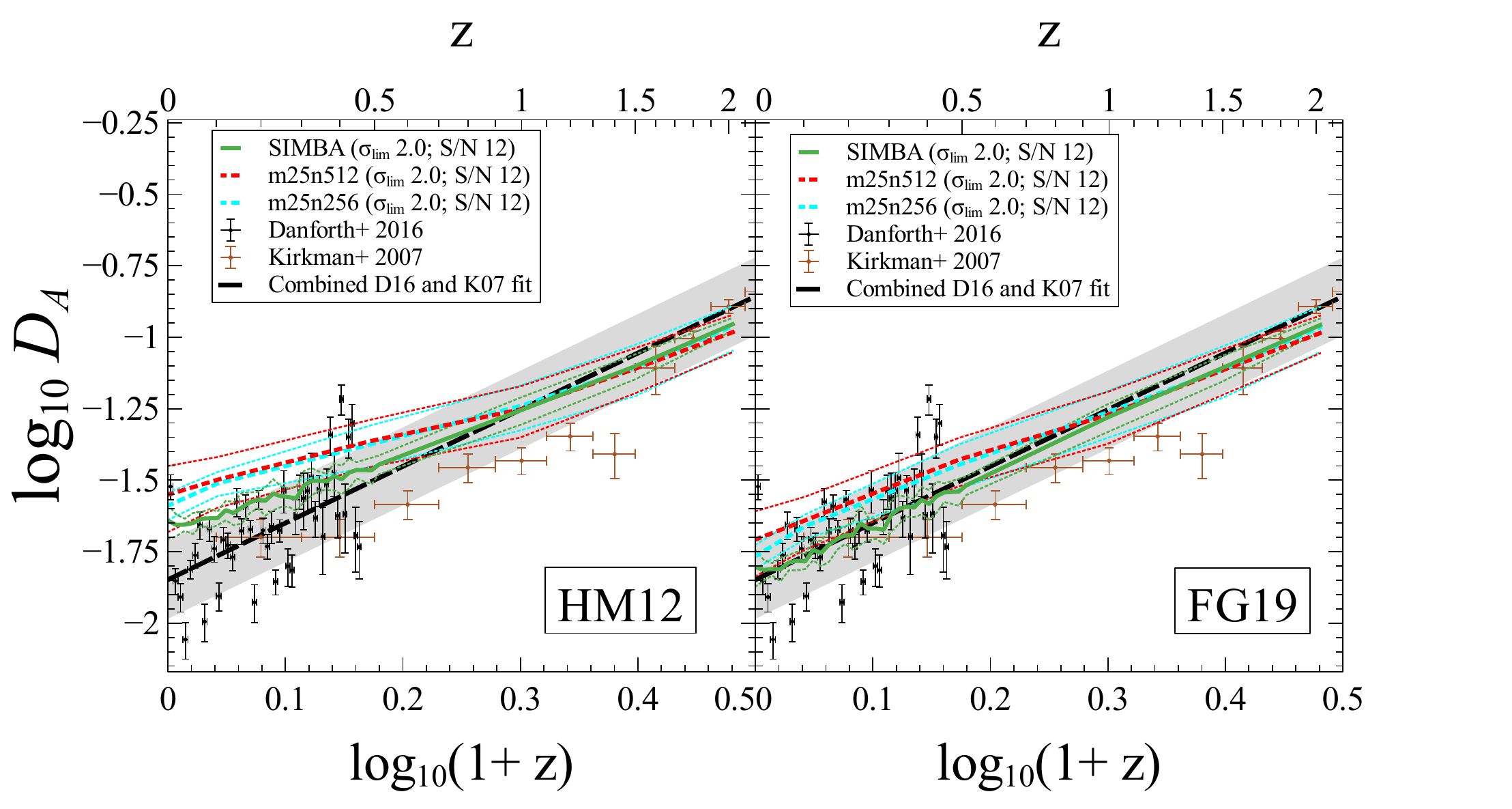}
    \vskip-0.2in
    \caption{\jacob{Graph showing effects of box size and resolution for 3 simulation runs of \simba\ varying these parameters, shown for both the HM12 (left) and FG19 (right) backgrounds. The green line shows the \simba\ run fiducial to this work, with a boxsize of $50 \hmpc$ and $512^3$ gas elements, labelled `\simba'; the red dashed line shows a run with a boxsize of $25 \hmpc$ and $512^3$ mass elements, labelled `m25n512'; and the cyan dashed line shows a run with a boxsize of $25 \hmpc$ and $256^3$ mass elements, labelled `m25n256'. The observational results from \citet{danforth-2016} and \citet{kirkman-2007}, along with a combined fit of their data are shown identically to previous graphs of $\DA$. } Among these tests, the change due to simulation volume is more important than numerical resolution, which has little impact. This owes to jet feedback being more prominent in massive galaxies that are underrepresented in a $25\hmpc$ volume.}
    \vskip-0.1in
\label{fig:m25-comparisons}
\end{figure*}
}

\section{Summary and Discussion}\label{sec:summary}

We have examined the evolution of the mean flux decrement in the \lya\ forest $\DA$ predicted in various simulations from the \simba\ suite, and used this to infer the \HI\ photo-ionisation rate as a function of redshift from $z=2\to 0$ by iteratively matching it to observations of $\DA$.  We consider the full \simba\ simulation that includes various forms of AGN feedback (jet, X-ray, and radiative), and compare it to identical simulations with either X-ray feedback or X-ray and jet feedback turned off.  We find greatest sensitivity to the inclusion of jet feedback:  With jet feedback turned off, we recover the so-called Photon Underproduction Crisis~\citep{kollmeier-2014} in which the \lya\ forest observations require $\gHI$ values at $z=0$ that are $\approx \times 6$ higher than inferred from source count modeling by \citet{haardt-madau-2012}.  Including jets (regardless of X-rays or radiative feedback), reduces this discrepancy to $\approx\times 2.5$, and further using an updated ionising background from \citet{faucher-2019} \romeel{now results in a reasonable match to $\DA$ both at $z=0$ and and its evolution since $z\sim 2$. } Hence in \simba, it appears that AGN jet feedback strongly mitigates the Photon Underproduction Crisis.

To understand the physical origin of the impact of jets, we examine the physical and evolutionary properties of the IGM, and their impact on $\DA$.  Our main findings are as follows:
\begin{itemize}
    \item Heating from AGN jets leads to a significantly increased fraction of baryons in the WHIM in \simba\ -- up to 70\% at $z=0$, versus 30\% when jets are excluded. This increase in the WHIM fraction comes primarily from a decrease in the baryon fraction in the diffuse IGM, from 39\% to 16\% at $z=0$, along with reductions in other cool phases.  With jets, the IGM baryon temperature distribution strongly peaks at $T\ga 10^6$K, instead of being predominantly at $T\la 10^5$K.
    \item The baryon fractions in various phases most strongly diverge between the full \simba\ and No-jet cases at $z\la 1$, when large black holes form that are responsible for quenching galaxies via jet heating. At $z\ga 1$, there is no PUC, as the predicted $\DA(z)$ in all \simba\ variants matches observations fairly well for either the HM12 or FG19 case.
    \item The decrease in the diffuse baryon fraction by $\times 2.5$ leads to a decrease in $\DA$ by a commensurate factor at $z=0$ in \simba\ with jet feedback.  Hence the main impact on $\DA$ of jet feedback is to remove IGM baryons from the \lya-absorbing phase via heating.  \romeelnew{This is corroborated by examining the pixel counts as a function of density, where \simba\ shows a new population of low-absorption fluxes near and above the cosmic mean density compared to No-jet, owing to jet heating.}
    \item  Assuming an FG19 background rather than HM12 results in the predicted $\DA$ matching observations over the full redshift range probed here ($z=0-2$), thus solving the PUC in \simba. Quantitatively, the photon underproduction factor $\fgam$ at $z=0$ is reduced by a factor of $\sim\times 2.5$ owing to the inclusion of jets, and by $\sim\times 2$ by using FG19 instead of HM12, thereby reducing $\fgam\approx 6\to 1.2$. 
    \item \romeel{The agreement in the redshift evolution of $\DA$ when including jets is a crucial success, as it highlights the importance of having a solution to the PUC that only impacts $\DA$ at late cosmic epochs, primarily at $z\la 1$.  This coincides with jet feedback becoming increasingly commonplace in order to yield today's red and dead galaxy population in \simba.}  
    \item \romeelnew{\simba\ further produces a good match to the flux probability distribution function (FPDF) as a function of redshift compared to COS GTO data, while No-jet significantly overproduces the FPDF.  This shows that \simba\ not only suppresses the mean absorption, but also the absorption as a function of flux in accord with data.}
    \item Examining a simulation with jets on but X-ray feedback off shows very similar results to the full \simba\ simulation with all AGN feedback modes on.  This demonstrates that it is the AGN jets that are responsible for heating the IGM and mitigating the PUC.
    \item Comparing \simba\ to \mufasa\ which used a different halo mass-based thermal quenching mechanism shows that \mufasa\ goes partways towards solving the PUC, but does not have as dramatic an effect as \simba's jets.  This suggests that careful measurements of $\gHI$ and $\DA$ could together provide constraints on AGN feedback mechanisms.
    \item \romeel{We have tested the sensitivity of these results to various numerical choices.  A key choice occurs in the continuum fitting process, and reasonable choices can lead to variations of up to $\sim 0.2$~dex in $\DA(z=0)$.  While this cannot fully explain the PUC, there is some uncertainty on the specific numbers quoted in above.  Generally, our choice of the key parameter $\siglim=2.0$ is conservative, in the sense that it results in less of a PUC than e.g. $\siglim=1.5$ used in \citet{danforth-2016}.  We further find that our results are insensitive to simulation resolution, but in \simba\ we do require a sufficient volume to representatively produce massive galaxies that drive jets.}
    %\item \simba\ jets make a lot more baryons end up in the WHIM
    %\item this reduces the amount of absorption in the diffuse phase, helping fix the PUC
    %\item mention that the jets fixing the PUC is quite impressive, considering that the AGN feedback model was not tuned to match absorption, but rather to match other features (quenched galaxies?). So, getting both right is quite promising
    %\item (suggest possible ways to compare this prediction to observation? Romeel was mentioning some different Oxygen ions (OVI and OVII I think?) - try to look at \citet{wijers-2019},  \citet{chen-2003}, and \citet{nicastro-2018}.
    %\item ``we also find better agreement with observed $\DA$ values when using the FG19 background versus the HM12 background''
    %\item AGN feedback models should be constrained by trying to replicate a number of observables: quenching of galaxies, hot halo gas fraction, and (we propose) low-redshift lyman-alpha $\DA$.
\end{itemize}

While some work has claimed that revising models of the ionising background is solely sufficient to solve the PUC, we find that it is also necessary for AGN jets to be modelled in order for the crisis to be fully resolved in \simba. AGN jets are phenomena that are known to exist, and it is heartening that their inclusion in state of the art simulations can also play a role in addressing other discrepancies between observation and theoretical predictions.

Our results broadly echo those presented in \citet{gurvich-2017}, who showed using the Illustris simulation that AGN feedback can have a strong impact on the diffuse \lya\ forest.  They likewise found that such heating, plus assuming an \citet{faucher-2009} background (which is slightly lower than FG19), essentially solved the PUC in Illustris. \simba\ has the advantage of a more plausible AGN feedback model that does a better job of quenching galaxies and does not over-evacuate hot halos, but the resulting impact on the IGM appears broadly comparable.  \romeel{Without jets, our results confirm every other fully hydrodynamical galaxy formation simulation's results that also find a PUC at the $\sim\times 4-6$ level.}\
%We are also able to investigate the specific impact of various AGN feedback modes, and we have investigated the redshift evolution of baryon phases and $\DA$ to clearly demonstrate that the impact of AGN jet feedback occurs mostly at $z\la 1$ and results in a strong reduction of diffuse \lya-absorbing gas in favor of WHIM gas.  % this is now in a bullet point...
\romeel{We also find that without jets, the redshift evolution of $\DA$ does not match observations.  Any model that aims to solve the $z=0$ PUC must also account for the fact that it is a late-time cosmic effect, essentially disappearing at $z\ga 1$.}

The large increase in WHIM baryon fraction should be testable with future observations, such as with high-ionisation oxygen absorption lines. The impact on \ion{O}{vi} absorption may be modest because in \simba\ the jet heating does not strongly increase the amount of $\sim 10^{5.5}$~K gas (Figure~\ref{fig:temperature-histograms-of-baryon-fractions}) where such absorption is strong, but rather moves gas to higher temperatures that would give rise to e.g. \ion{O}{vii} absorption in the soft X-rays\jacob{, or \ion{Ne}{viii} in the extreme UV.}\ Current constraints are insufficient to discriminate between our jet vs. no-jet predictions, but upcoming facilities such as {\it Athena} and {\it Lynx} would be ideal for this.  Another potential avenue for constraints is examining Sunyaev-Zel'dovich integrated IGM pressure measurements~\citep[e.g.][]{lim-2018,degraaff-2019}, which could provide constraints on the phase space distribution of IGM baryons.  We plan to investigate whether \simba\ satisfies these constraints in future work.

The shape of the \HI\ column density distribution is also an important constraint for solving the PUC.  We have sidestepped this issue here, even though it was an important consideration in previous works~\citep{kollmeier-2014,shull-2015,gurvich-2017}. Any solution to the PUC must also impact the column density distribution in a way that remains concordant with observations.  A proper comparison of this, however, requires carefully mimicking the observational signal to noise, line spread function, wavelength coverage, and profile fitting algorithm used for the data.  It is worth noting that in \citet{dave-2001b}, the observed column density distribution using high-resolution HST/STIS data was found to be significantly steeper than that found by \citet{danforth-2016} using lower resolution HST/COS data, illustrating this sensitivity.  We plan to conduct side-by-side Voigt profile fitting comparisons of absorber statistics in the future, but the PUC is already evident even when considering the stacked statistic of the mean flux decrement.

Broadly, our conclusions highlight the perhaps surprising point that the ionisation level of the low-redshift IGM as traced by \lya\ absorption can potentially be strongly impacted by AGN feedback originating deep within massive galaxies.  While current uncertainties around determining the low-$z$ metagalactic photo-ionisation rate complicate the interpretation, this nonetheless provides new avenues to constrain AGN feedback models in a regime far removed from where it is typically constrained via the properties of quenched galaxies and their black holes.

\section*{Acknowledgements}

The authors thank Sarah Appleby, Blakesley Burkhart, Neal Katz, and Juna Kollmeier for helpful discussions.
We thank the {\sc yt} team for development and support of {\sc yt}, and Bernhard R\"ottgers and Miha Cernetic for development of {\sc pygad}.
RD acknowledges support from the Wolfson Research Merit Award program of the U.K. Royal Society. 
DS was supported by the European Research Council, under grant no. 670193.
DAA acknowledges support by the Flatiron Institute, which is supported by the Simons Foundation
This work used the DiRAC@Durham facility managed by the Institute for Computational Cosmology on behalf of the STFC DiRAC HPC Facility. The equipment was funded by BEIS capital funding via STFC capital grants ST/P002293/1, ST/R002371/1 and ST/S002502/1, Durham University and STFC operations grant ST/R000832/1. DiRAC is part of the National e-Infrastructure.
%%TC:endignore

\section*{Data availability}

The simulation data underlying this article are available at {\tt simba.roe.ac.uk}. The derived data generated in this research will be shared on reasonable request to the corresponding author.

%%%%%%%%%%%%%%%%%%%%%%%%%%%%%%%%%%%%%%%%%%%%%%%%%%



%\bibliographystyle{mnras}
\bibliographystyle{mnras}
\bibliography{puc-paper}

\begin{thebibliography}{}
\makeatletter
\relax
\def\mn@urlcharsother{\let\do\@makeother \do\$\do\&\do\#\do\^\do\_\do\%\do\~}
\def\mn@doi{\begingroup\mn@urlcharsother \@ifnextchar [ {\mn@doi@}
  {\mn@doi@[]}}
\def\mn@doi@[#1]#2{\def\@tempa{#1}\ifx\@tempa\@empty \href
  {http://dx.doi.org/#2} {doi:#2}\else \href {http://dx.doi.org/#2} {#1}\fi
  \endgroup}
\def\mn@eprint#1#2{\mn@eprint@#1:#2::\@nil}
\def\mn@eprint@arXiv#1{\href {http://arxiv.org/abs/#1} {{\tt arXiv:#1}}}
\def\mn@eprint@dblp#1{\href {http://dblp.uni-trier.de/rec/bibtex/#1.xml}
  {dblp:#1}}
\def\mn@eprint@#1:#2:#3:#4\@nil{\def\@tempa {#1}\def\@tempb {#2}\def\@tempc
  {#3}\ifx \@tempc \@empty \let \@tempc \@tempb \let \@tempb \@tempa \fi \ifx
  \@tempb \@empty \def\@tempb {arXiv}\fi \@ifundefined
  {mn@eprint@\@tempb}{\@tempb:\@tempc}{\expandafter \expandafter \csname
  mn@eprint@\@tempb\endcsname \expandafter{\@tempc}}}

\bibitem[\protect\citeauthoryear{{Abel}, {Anninos}, {Zhang}  \&
  {Norman}}{{Abel} et~al.}{1997}]{abel-1997}
{Abel} T.,  {Anninos} P.,  {Zhang} Y.,   {Norman} M.~L.,  1997, \mn@doi [\na]
  {10.1016/S1384-1076(97)00010-9}, \href
  {https://ui.adsabs.harvard.edu/abs/1997NewA....2..181A} {2, 181}

\bibitem[\protect\citeauthoryear{{Angl{\'e}s-Alc{\'a}zar}, {{\"O}zel}  \&
  {Dav{\'e}}}{{Angl{\'e}s-Alc{\'a}zar} et~al.}{2013}]{angles-alcazar-2013}
{Angl{\'e}s-Alc{\'a}zar} D.,  {{\"O}zel} F.,   {Dav{\'e}} R.,  2013, \mn@doi
  [\apj] {10.1088/0004-637X/770/1/5}, \href
  {https://ui.adsabs.harvard.edu/abs/2013ApJ...770....5A} {770, 5}

\bibitem[\protect\citeauthoryear{{Angl{\'e}s-Alc{\'a}zar}, {Dav{\'e}},
  {Faucher-Gigu{\`e}re}, {{\"O}zel}  \& {Hopkins}}{{Angl{\'e}s-Alc{\'a}zar}
  et~al.}{2017a}]{angles-alcazar-2017a}
{Angl{\'e}s-Alc{\'a}zar} D.,  {Dav{\'e}} R.,  {Faucher-Gigu{\`e}re} C.-A.,
  {{\"O}zel} F.,   {Hopkins} P.~F.,  2017a, \mn@doi [\mnras]
  {10.1093/mnras/stw2565}, \href
  {https://ui.adsabs.harvard.edu/abs/2017MNRAS.464.2840A} {464, 2840}

\bibitem[\protect\citeauthoryear{{Angl{\'e}s-Alc{\'a}zar},
  {Faucher-Gigu{\`e}re}, {Kere{\v{s}}}, {Hopkins}, {Quataert}  \&
  {Murray}}{{Angl{\'e}s-Alc{\'a}zar} et~al.}{2017b}]{angles-alcazar-2017b}
{Angl{\'e}s-Alc{\'a}zar} D.,  {Faucher-Gigu{\`e}re} C.-A.,  {Kere{\v{s}}} D.,
  {Hopkins} P.~F.,  {Quataert} E.,   {Murray} N.,  2017b, \mn@doi [Monthly
  Notices of the Royal Astronomical Society] {10.1093/mnras/stx1517}, \href
  {https://ui.adsabs.harvard.edu/abs/2017MNRAS.470.4698A} {470, 4698}

\bibitem[\protect\citeauthoryear{{Best} \& {Heckman}}{{Best} \&
  {Heckman}}{2012}]{best-2012}
{Best} P.~N.,  {Heckman} T.~M.,  2012, \mn@doi [\mnras]
  {10.1111/j.1365-2966.2012.20414.x}, \href
  {https://ui.adsabs.harvard.edu/abs/2012MNRAS.421.1569B} {421, 1569}

\bibitem[\protect\citeauthoryear{{Bolton}, {Puchwein}, {Sijacki}, {Haehnelt},
  {Kim}, {Meiksin}, {Regan}  \& {Viel}}{{Bolton} et~al.}{2017}]{bolton-2017}
{Bolton} J.~S.,  {Puchwein} E.,  {Sijacki} D.,  {Haehnelt} M.~G.,  {Kim} T.-S.,
   {Meiksin} A.,  {Regan} J.~A.,   {Viel} M.,  2017, \mn@doi [\mnras]
  {10.1093/mnras/stw2397}, \href
  {https://ui.adsabs.harvard.edu/abs/2017MNRAS.464..897B} {464, 897}

\bibitem[\protect\citeauthoryear{{Bondi}}{{Bondi}}{1952}]{bondi-1952}
{Bondi} H.,  1952, \mn@doi [\mnras] {10.1093/mnras/112.2.195}, \href
  {https://ui.adsabs.harvard.edu/abs/1952MNRAS.112..195B} {112, 195}

\bibitem[\protect\citeauthoryear{{Borrow}, {Angles-Alcazar}  \&
  {Dave}}{{Borrow} et~al.}{2019}]{borrow-2019}
{Borrow} J.,  {Angles-Alcazar} D.,   {Dave} R.,  2019, arXiv e-prints, \href
  {https://ui.adsabs.harvard.edu/abs/2019arXiv191000594B} {p. arXiv:1910.00594}

\bibitem[\protect\citeauthoryear{{Broderick}, {Chang}  \&
  {Pfrommer}}{{Broderick} et~al.}{2012}]{broderick-2012}
{Broderick} A.~E.,  {Chang} P.,   {Pfrommer} C.,  2012, \mn@doi [\apj]
  {10.1088/0004-637X/752/1/22}, \href
  {https://ui.adsabs.harvard.edu/abs/2012ApJ...752...22B} {752, 22}

\bibitem[\protect\citeauthoryear{{Burchett} et~al.,}{{Burchett}
  et~al.}{2019}]{burchett-2019}
{Burchett} J.~N.,  et~al., 2019, \mn@doi [\apjl] {10.3847/2041-8213/ab1f7f},
  \href {https://ui.adsabs.harvard.edu/abs/2019ApJ...877L..20B} {877, L20}

\bibitem[\protect\citeauthoryear{{Cen} \& {Ostriker}}{{Cen} \&
  {Ostriker}}{1999}]{cen-ostriker-1999}
{Cen} R.,  {Ostriker} J.~P.,  1999, \mn@doi [\apj] {10.1086/306949}, \href
  {https://ui.adsabs.harvard.edu/abs/1999ApJ...514....1C} {514, 1}

\bibitem[\protect\citeauthoryear{{Chen}, {Weinberg}, {Katz}  \&
  {Dav{\'e}}}{{Chen} et~al.}{2003}]{chen-2003}
{Chen} X.,  {Weinberg} D.~H.,  {Katz} N.,   {Dav{\'e}} R.,  2003, \mn@doi
  [\apj] {10.1086/376751}, \href
  {https://ui.adsabs.harvard.edu/abs/2003ApJ...594...42C} {594, 42}

\bibitem[\protect\citeauthoryear{{Choi}, {Ostriker}, {Naab}  \&
  {Johansson}}{{Choi} et~al.}{2012}]{choi-2012}
{Choi} E.,  {Ostriker} J.~P.,  {Naab} T.,   {Johansson} P.~H.,  2012, \mn@doi
  [\apj] {10.1088/0004-637X/754/2/125}, \href
  {https://ui.adsabs.harvard.edu/abs/2012ApJ...754..125C} {754, 125}

\bibitem[\protect\citeauthoryear{Danforth et~al.,}{Danforth
  et~al.}{2016}]{danforth-2016}
Danforth C.~W.,  et~al., 2016, \mn@doi [The Astrophysical Journal]
  {10.3847/0004-637x/817/2/111}, 817, 111

\bibitem[\protect\citeauthoryear{Dave \& Tripp}{Dave \&
  Tripp}{2001}]{dave-2001b}
Dave R.,  Tripp T.~M.,  2001, \mn@doi [The Astrophysical Journal]
  {10.1086/320977}, 553, 528

\bibitem[\protect\citeauthoryear{{Dav{\'e}}, {Hernquist}, {Katz}  \&
  {Weinberg}}{{Dav{\'e}} et~al.}{1999}]{dave-1999}
{Dav{\'e}} R.,  {Hernquist} L.,  {Katz} N.,   {Weinberg} D.~H.,  1999, \mn@doi
  [\apj] {10.1086/306722}, \href
  {https://ui.adsabs.harvard.edu/\#abs/1999ApJ...511..521D} {511, 521}

\bibitem[\protect\citeauthoryear{Dave et~al.,}{Dave et~al.}{2001}]{dave-2001a}
Dave R.,  et~al., 2001, \mn@doi [The Astrophysical Journal] {10.1086/320548},
  552, 473

\bibitem[\protect\citeauthoryear{{Dav{\'e}}, {Oppenheimer}  \& {Sivanand
  am}}{{Dav{\'e}} et~al.}{2008}]{dave-2008}
{Dav{\'e}} R.,  {Oppenheimer} B.~D.,   {Sivanand am} S.,  2008, \mn@doi
  [\mnras] {10.1111/j.1365-2966.2008.13906.x}, \href
  {https://ui.adsabs.harvard.edu/abs/2008MNRAS.391..110D} {391, 110}

\bibitem[\protect\citeauthoryear{{Dav{\'e}}, {Oppenheimer}, {Katz}, {Kollmeier}
   \& {Weinberg}}{{Dav{\'e}} et~al.}{2010}]{dave-2010}
{Dav{\'e}} R.,  {Oppenheimer} B.~D.,  {Katz} N.,  {Kollmeier} J.~A.,
  {Weinberg} D.~H.,  2010, \mn@doi [\mnras] {10.1111/j.1365-2966.2010.17279.x},
  \href {https://ui.adsabs.harvard.edu/\#abs/2010MNRAS.408.2051D} {408, 2051}

\bibitem[\protect\citeauthoryear{{Dav{\'e}}, {Katz}, {Oppenheimer}, {Kollmeier}
   \& {Weinberg}}{{Dav{\'e}} et~al.}{2013}]{dave-2013}
{Dav{\'e}} R.,  {Katz} N.,  {Oppenheimer} B.~D.,  {Kollmeier} J.~A.,
  {Weinberg} D.~H.,  2013, \mn@doi [\mnras] {10.1093/mnras/stt1274}, \href
  {https://ui.adsabs.harvard.edu/abs/2013MNRAS.434.2645D} {434, 2645}

\bibitem[\protect\citeauthoryear{{Dav{\'e}}, {Thompson}  \&
  {Hopkins}}{{Dav{\'e}} et~al.}{2016}]{dave-2016}
{Dav{\'e}} R.,  {Thompson} R.,   {Hopkins} P.~F.,  2016, \mn@doi [\mnras]
  {10.1093/mnras/stw1862}, \href
  {https://ui.adsabs.harvard.edu/abs/2016MNRAS.462.3265D} {462, 3265}

\bibitem[\protect\citeauthoryear{{Dav{\'e}}, {Rafieferantsoa}  \&
  {Thompson}}{{Dav{\'e}} et~al.}{2017}]{dave-2017b}
{Dav{\'e}} R.,  {Rafieferantsoa} M.~H.,   {Thompson} R.~J.,  2017, \mn@doi
  [\mnras] {10.1093/mnras/stx1693}, \href
  {https://ui.adsabs.harvard.edu/abs/2017MNRAS.471.1671D} {471, 1671}

\bibitem[\protect\citeauthoryear{{Dav{\'e}}, {Angl{\'e}s-Alc{\'a}zar},
  {Narayanan}, {Li}, {Rafieferantsoa}  \& {Appleby}}{{Dav{\'e}}
  et~al.}{2019}]{dave-2019}
{Dav{\'e}} R.,  {Angl{\'e}s-Alc{\'a}zar} D.,  {Narayanan} D.,  {Li} Q.,
  {Rafieferantsoa} M.~H.,   {Appleby} S.,  2019, \mn@doi [\mnras]
  {10.1093/mnras/stz937}, \href
  {https://ui.adsabs.harvard.edu/abs/2019MNRAS.486.2827D} {486, 2827}

\bibitem[\protect\citeauthoryear{{Fabian}}{{Fabian}}{2012}]{fabian-2012}
{Fabian} A.~C.,  2012, \mn@doi [\araa] {10.1146/annurev-astro-081811-125521},
  \href {https://ui.adsabs.harvard.edu/abs/2012ARA&A..50..455F} {50, 455}

\bibitem[\protect\citeauthoryear{{Faucher-Gigu{\`e}re}}{{Faucher-Gigu{\`e}re}}{2019}]{faucher-2019}
{Faucher-Gigu{\`e}re} C.-A.,  2019, arXiv e-prints, \href
  {https://ui.adsabs.harvard.edu/abs/2019arXiv190308657F} {p. arXiv:1903.08657}

\bibitem[\protect\citeauthoryear{{Faucher-Gigu{\`e}re}, {Lidz}, {Zaldarriaga}
  \& {Hernquist}}{{Faucher-Gigu{\`e}re} et~al.}{2009}]{faucher-2009}
{Faucher-Gigu{\`e}re} C.-A.,  {Lidz} A.,  {Zaldarriaga} M.,   {Hernquist} L.,
  2009, \mn@doi [\apj] {10.1088/0004-637X/703/2/1416}, \href
  {https://ui.adsabs.harvard.edu/abs/2009ApJ...703.1416F} {703, 1416}

\bibitem[\protect\citeauthoryear{{Fumagalli}, {Haardt}, {Theuns}, {Morris},
  {Cantalupo}, {Madau}  \& {Fossati}}{{Fumagalli}
  et~al.}{2017}]{fumagalli-2017}
{Fumagalli} M.,  {Haardt} F.,  {Theuns} T.,  {Morris} S.~L.,  {Cantalupo} S.,
  {Madau} P.,   {Fossati} M.,  2017, \mn@doi [\mnras] {10.1093/mnras/stx398},
  \href {https://ui.adsabs.harvard.edu/abs/2017MNRAS.467.4802F} {467, 4802}

\bibitem[\protect\citeauthoryear{{Gaikwad}, {Khaire}, {Choudhury}  \&
  {Srianand}}{{Gaikwad} et~al.}{2017a}]{gaikwad-2017a}
{Gaikwad} P.,  {Khaire} V.,  {Choudhury} T.~R.,   {Srianand} R.,  2017a,
  \mn@doi [\mnras] {10.1093/mnras/stw3086}, \href
  {https://ui.adsabs.harvard.edu/abs/2017MNRAS.466..838G} {466, 838}

\bibitem[\protect\citeauthoryear{{Gaikwad}, {Srianand}, {Choudhury}  \&
  {Khaire}}{{Gaikwad} et~al.}{2017b}]{gaikwad-2017b}
{Gaikwad} P.,  {Srianand} R.,  {Choudhury} T.~R.,   {Khaire} V.,  2017b,
  \mn@doi [\mnras] {10.1093/mnras/stx248}, \href
  {https://ui.adsabs.harvard.edu/abs/2017MNRAS.467.3172G} {467, 3172}

\bibitem[\protect\citeauthoryear{{Genel} et~al.,}{{Genel}
  et~al.}{2014}]{genel-2014}
{Genel} S.,  et~al., 2014, \mn@doi [\mnras] {10.1093/mnras/stu1654}, \href
  {https://ui.adsabs.harvard.edu/abs/2014MNRAS.445..175G} {445, 175}

\bibitem[\protect\citeauthoryear{{Gurvich}, {Burkhart}  \& {Bird}}{{Gurvich}
  et~al.}{2017}]{gurvich-2017}
{Gurvich} A.,  {Burkhart} B.,   {Bird} S.,  2017, \mn@doi [The Astrophysical
  Journal] {10.3847/1538-4357/835/2/175}, \href
  {https://ui.adsabs.harvard.edu/\#abs/2017ApJ...835..175G} {835, 175}

\bibitem[\protect\citeauthoryear{{Haardt} \& {Madau}}{{Haardt} \&
  {Madau}}{1996}]{haardt-madau-1996}
{Haardt} F.,  {Madau} P.,  1996, \mn@doi [\apj] {10.1086/177035}, \href
  {https://ui.adsabs.harvard.edu/abs/1996ApJ...461...20H} {461, 20}

\bibitem[\protect\citeauthoryear{{Haardt} \& {Madau}}{{Haardt} \&
  {Madau}}{2001}]{haardt-madau-2001}
{Haardt} F.,  {Madau} P.,  2001, in {Neumann} D.~M.,  {Tran} J.~T.~V.,  eds,
  Clusters of Galaxies and the High Redshift Universe Observed in X-rays. p.~64
  (\mn@eprint {} {astro-ph/0106018})

\bibitem[\protect\citeauthoryear{{Haardt} \& {Madau}}{{Haardt} \&
  {Madau}}{2012}]{haardt-madau-2012}
{Haardt} F.,  {Madau} P.,  2012, \mn@doi [\apj] {10.1088/0004-637X/746/2/125},
  \href {https://ui.adsabs.harvard.edu/abs/2012ApJ...746..125H} {746, 125}

\bibitem[\protect\citeauthoryear{{Heckman} \& {Best}}{{Heckman} \&
  {Best}}{2014}]{heckman-2014}
{Heckman} T.~M.,  {Best} P.~N.,  2014, \mn@doi [\araa]
  {10.1146/annurev-astro-081913-035722}, \href
  {https://ui.adsabs.harvard.edu/abs/2014ARA&A..52..589H} {52, 589}

\bibitem[\protect\citeauthoryear{{Henden}, {Puchwein}, {Shen}  \&
  {Sijacki}}{{Henden} et~al.}{2018}]{henden-2018}
{Henden} N.~A.,  {Puchwein} E.,  {Shen} S.,   {Sijacki} D.,  2018, \mn@doi
  [\mnras] {10.1093/mnras/sty1780}, \href
  {https://ui.adsabs.harvard.edu/abs/2018MNRAS.479.5385H} {479, 5385}

\bibitem[\protect\citeauthoryear{{Hopkins}}{{Hopkins}}{2015}]{hopkins-2015}
{Hopkins} P.~F.,  2015, \mn@doi [\mnras] {10.1093/mnras/stv195}, \href
  {https://ui.adsabs.harvard.edu/\#abs/2015MNRAS.450...53H} {450, 53}

\bibitem[\protect\citeauthoryear{{Hopkins} \& {Quataert}}{{Hopkins} \&
  {Quataert}}{2011}]{hopkins-2011}
{Hopkins} P.~F.,  {Quataert} E.,  2011, \mn@doi [\mnras]
  {10.1111/j.1365-2966.2011.18542.x}, \href
  {https://ui.adsabs.harvard.edu/abs/2011MNRAS.415.1027H} {415, 1027}

\bibitem[\protect\citeauthoryear{{Hui} \& {Gnedin}}{{Hui} \&
  {Gnedin}}{1997}]{hui-gnedin-1997}
{Hui} L.,  {Gnedin} N.~Y.,  1997, \mn@doi [\mnras] {10.1093/mnras/292.1.27},
  \href {https://ui.adsabs.harvard.edu/abs/1997MNRAS.292...27H} {292, 27}

\bibitem[\protect\citeauthoryear{{Katz}, {Weinberg}  \& {Hernquist}}{{Katz}
  et~al.}{1996}]{katz-1996}
{Katz} N.,  {Weinberg} D.~H.,   {Hernquist} L.,  1996, \mn@doi [\apjs]
  {10.1086/192305}, \href
  {https://ui.adsabs.harvard.edu/abs/1996ApJS..105...19K} {105, 19}

\bibitem[\protect\citeauthoryear{{Kennicutt}}{{Kennicutt}}{1998}]{kennicutt-1998}
{Kennicutt} Robert~C. J.,  1998, \mn@doi [\araa]
  {10.1146/annurev.astro.36.1.189}, \href
  {https://ui.adsabs.harvard.edu/abs/1998ARA&A..36..189K} {36, 189}

\bibitem[\protect\citeauthoryear{Khaire \& Srianand}{Khaire \&
  Srianand}{2015}]{khaire-2015}
Khaire V.,  Srianand R.,  2015, \mn@doi [Monthly Notices of the Royal
  Astronomical Society: Letters] {10.1093/mnrasl/slv060}, 451, L30

\bibitem[\protect\citeauthoryear{{Khaire} et~al.,}{{Khaire}
  et~al.}{2019}]{khaire-2019}
{Khaire} V.,  et~al., 2019, \mn@doi [\mnras] {10.1093/mnras/stz344}, \href
  {https://ui.adsabs.harvard.edu/abs/2019MNRAS.486..769K} {486, 769}

\bibitem[\protect\citeauthoryear{{Kirkman}, {Tytler}, {Lubin}  \&
  {Charlton}}{{Kirkman} et~al.}{2007}]{kirkman-2007}
{Kirkman} D.,  {Tytler} D.,  {Lubin} D.,   {Charlton} J.,  2007, \mn@doi
  [\mnras] {10.1111/j.1365-2966.2007.11502.x}, \href
  {https://ui.adsabs.harvard.edu/abs/2007MNRAS.376.1227K} {376, 1227}

\bibitem[\protect\citeauthoryear{{Kitayama} \& {Suto}}{{Kitayama} \&
  {Suto}}{1996}]{kitayama-1996}
{Kitayama} T.,  {Suto} Y.,  1996, \mn@doi [\apj] {10.1086/177797}, \href
  {https://ui.adsabs.harvard.edu/abs/1996ApJ...469..480K} {469, 480}

\bibitem[\protect\citeauthoryear{{Kollmeier} et~al.,}{{Kollmeier}
  et~al.}{2014}]{kollmeier-2014}
{Kollmeier} J.~A.,  et~al., 2014, \mn@doi [\apjl]
  {10.1088/2041-8205/789/2/L32}, \href
  {https://ui.adsabs.harvard.edu/abs/2014ApJ...789L..32K} {789, L32}

\bibitem[\protect\citeauthoryear{{Kormendy} \& {Ho}}{{Kormendy} \&
  {Ho}}{2013}]{kormendy-2013}
{Kormendy} J.,  {Ho} L.~C.,  2013, \mn@doi [\araa]
  {10.1146/annurev-astro-082708-101811}, \href
  {https://ui.adsabs.harvard.edu/abs/2013ARA&A..51..511K} {51, 511}

\bibitem[\protect\citeauthoryear{{Krumholz} \& {Gnedin}}{{Krumholz} \&
  {Gnedin}}{2011}]{krumholz-2011}
{Krumholz} M.~R.,  {Gnedin} N.~Y.,  2011, \mn@doi [\apj]
  {10.1088/0004-637X/729/1/36}, \href
  {https://ui.adsabs.harvard.edu/abs/2011ApJ...729...36K} {729, 36}

\bibitem[\protect\citeauthoryear{{Kulkarni}, {Worseck}  \&
  {Hennawi}}{{Kulkarni} et~al.}{2019}]{kulkarni-2019}
{Kulkarni} G.,  {Worseck} G.,   {Hennawi} J.~F.,  2019, \mn@doi [\mnras]
  {10.1093/mnras/stz1493}, \href
  {https://ui.adsabs.harvard.edu/abs/2019MNRAS.488.1035K} {488, 1035}

\bibitem[\protect\citeauthoryear{{Li}, {Narayanan}  \& {Dav{\'e}}}{{Li}
  et~al.}{2019}]{li-2019}
{Li} Q.,  {Narayanan} D.,   {Dav{\'e}} R.,  2019, \mn@doi [\mnras]
  {10.1093/mnras/stz2684}, \href
  {https://ui.adsabs.harvard.edu/abs/2019MNRAS.490.1425L} {490, 1425}

\bibitem[\protect\citeauthoryear{{Lim}, {Mo}, {Wang}  \& {Yang}}{{Lim}
  et~al.}{2018}]{lim-2018}
{Lim} S.~H.,  {Mo} H.~J.,  {Wang} H.,   {Yang} X.,  2018, \mn@doi [\mnras]
  {10.1093/mnras/sty2126}, \href
  {https://ui.adsabs.harvard.edu/abs/2018MNRAS.480.4017L} {480, 4017}

\bibitem[\protect\citeauthoryear{Meiksin}{Meiksin}{2009}]{meiksin-2009}
Meiksin A.~A.,  2009, \mn@doi [Rev. Mod. Phys.] {10.1103/RevModPhys.81.1405},
  81, 1405

\bibitem[\protect\citeauthoryear{{Nicastro} et~al.,}{{Nicastro}
  et~al.}{2018}]{nicastro-2018}
{Nicastro} F.,  et~al., 2018, \mn@doi [\nat] {10.1038/s41586-018-0204-1}, \href
  {https://ui.adsabs.harvard.edu/abs/2018Natur.558..406N} {558, 406}

\bibitem[\protect\citeauthoryear{{Oppenheimer} \& {Dav{\'e}}}{{Oppenheimer} \&
  {Dav{\'e}}}{2006}]{oppenheimer-2006}
{Oppenheimer} B.~D.,  {Dav{\'e}} R.,  2006, \mn@doi [\mnras]
  {10.1111/j.1365-2966.2006.10989.x}, \href
  {https://ui.adsabs.harvard.edu/abs/2006MNRAS.373.1265O} {373, 1265}

\bibitem[\protect\citeauthoryear{{Perna}, {Lanzuisi}, {Brusa}, {Mignoli}  \&
  {Cresci}}{{Perna} et~al.}{2017}]{perna-2017a}
{Perna} M.,  {Lanzuisi} G.,  {Brusa} M.,  {Mignoli} M.,   {Cresci} G.,  2017,
  \mn@doi [\aap] {10.1051/0004-6361/201630369}, \href
  {https://ui.adsabs.harvard.edu/abs/2017A&A...603A..99P} {603, A99}

\bibitem[\protect\citeauthoryear{{Planck Collaboration} et~al.,}{{Planck
  Collaboration} et~al.}{2016}]{planck-2016}
{Planck Collaboration} et~al., 2016, \mn@doi [A\&A]
  {10.1051/0004-6361/201525830}, 594, A13

\bibitem[\protect\citeauthoryear{{Rahmati}, {Pawlik}, {Rai{\v{c}}evi{\'c}}  \&
  {Schaye}}{{Rahmati} et~al.}{2013}]{rahmati-2013}
{Rahmati} A.,  {Pawlik} A.~H.,  {Rai{\v{c}}evi{\'c}} M.,   {Schaye} J.,  2013,
  \mn@doi [Monthly Notices of the Royal Astronomical Society]
  {10.1093/mnras/stt066}, \href
  {https://ui.adsabs.harvard.edu/abs/2013MNRAS.430.2427R} {430, 2427}

\bibitem[\protect\citeauthoryear{{Rauch} et~al.,}{{Rauch}
  et~al.}{1997}]{rauch-1997}
{Rauch} M.,  et~al., 1997, \mn@doi [The Astrophysical Journal]
  {10.1086/304765}, \href
  {https://ui.adsabs.harvard.edu/abs/1997ApJ...489....7R} {489, 7}

\bibitem[\protect\citeauthoryear{{Robson} \& {Dav{\'e}}}{{Robson} \&
  {Dav{\'e}}}{2020}]{robson-2020}
{Robson} D.,  {Dav{\'e}} R.,  2020, arXiv e-prints, \href
  {https://ui.adsabs.harvard.edu/abs/2020arXiv200304115R} {p. arXiv:2003.04115}

\bibitem[\protect\citeauthoryear{{R{\"o}ttgers}, {Naab}, {Cernetic},
  {Dav{\'e}}, {Kauffmann}, {Borthakur}  \& {Foidl}}{{R{\"o}ttgers}
  et~al.}{2020}]{roettgers-2020}
{R{\"o}ttgers} B.,  {Naab} T.,  {Cernetic} M.,  {Dav{\'e}} R.,  {Kauffmann} G.,
   {Borthakur} S.,   {Foidl} H.,  2020, \mn@doi [\mnras]
  {10.1093/mnras/staa1490}, \href
  {https://ui.adsabs.harvard.edu/abs/2020MNRAS.496..152R} {496, 152}

\bibitem[\protect\citeauthoryear{{Rutkowski} et~al.,}{{Rutkowski}
  et~al.}{2016}]{rutkowski-2016}
{Rutkowski} M.~J.,  et~al., 2016, \mn@doi [\apj] {10.3847/0004-637X/819/1/81},
  \href {https://ui.adsabs.harvard.edu/abs/2016ApJ...819...81R} {819, 81}

\bibitem[\protect\citeauthoryear{{Schaye} et~al.,}{{Schaye}
  et~al.}{2015}]{schaye-2015}
{Schaye} J.,  et~al., 2015, \mn@doi [\mnras] {10.1093/mnras/stu2058}, \href
  {https://ui.adsabs.harvard.edu/abs/2015MNRAS.446..521S} {446, 521}

\bibitem[\protect\citeauthoryear{Shull, Smith  \& Danforth}{Shull
  et~al.}{2012}]{shull-2012}
Shull J.~M.,  Smith B.~D.,   Danforth C.~W.,  2012, \mn@doi [The Astrophysical
  Journal] {10.1088/0004-637x/759/1/23}, 759, 23

\bibitem[\protect\citeauthoryear{Shull, Moloney, Danforth  \& Tilton}{Shull
  et~al.}{2015}]{shull-2015}
Shull J.~M.,  Moloney J.,  Danforth C.~W.,   Tilton E.~M.,  2015, \mn@doi [The
  Astrophysical Journal] {10.1088/0004-637x/811/1/3}, 811, 3

\bibitem[\protect\citeauthoryear{{Smith}, {Hallman}, {Shull}  \&
  {O'Shea}}{{Smith} et~al.}{2011}]{smith-2011}
{Smith} B.~D.,  {Hallman} E.~J.,  {Shull} J.~M.,   {O'Shea} B.~W.,  2011,
  \mn@doi [\apj] {10.1088/0004-637X/731/1/6}, \href
  {https://ui.adsabs.harvard.edu/abs/2011ApJ...731....6S} {731, 6}

\bibitem[\protect\citeauthoryear{{Smith} et~al.,}{{Smith}
  et~al.}{2017}]{smith-2017}
{Smith} B.~D.,  et~al., 2017, \mn@doi [\mnras] {10.1093/mnras/stw3291}, \href
  {https://ui.adsabs.harvard.edu/abs/2017MNRAS.466.2217S} {466, 2217}

\bibitem[\protect\citeauthoryear{{Somerville} \& {Dav{\'e}}}{{Somerville} \&
  {Dav{\'e}}}{2015}]{somerville-2015}
{Somerville} R.~S.,  {Dav{\'e}} R.,  2015, \mn@doi [\araa]
  {10.1146/annurev-astro-082812-140951}, \href
  {https://ui.adsabs.harvard.edu/abs/2015ARA&A..53...51S} {53, 51}

\bibitem[\protect\citeauthoryear{{Sorini}}{{Sorini}}{2017}]{Sorini-phd}
{Sorini} D.,  2017, PhD thesis, International Max Planck Research School for
  Astronomy and Cosmic Physics at the University of Heidelberg (IMPRS-HD),
  Germany

\bibitem[\protect\citeauthoryear{{Sorini}, {O{\~n}orbe}, {Hennawi}  \&
  {Luki{\'c}}}{{Sorini} et~al.}{2018}]{sorini-2018}
{Sorini} D.,  {O{\~n}orbe} J.,  {Hennawi} J.~F.,   {Luki{\'c}} Z.,  2018,
  \mn@doi [\apj] {10.3847/1538-4357/aabb52}, \href
  {https://ui.adsabs.harvard.edu/abs/2018ApJ...859..125S} {859, 125}

\bibitem[\protect\citeauthoryear{{Thomas}, {Dav{\'e}}, {Angl{\'e}s-Alc{\'a}zar}
   \& {Jarvis}}{{Thomas} et~al.}{2019}]{thomas-2019}
{Thomas} N.,  {Dav{\'e}} R.,  {Angl{\'e}s-Alc{\'a}zar} D.,   {Jarvis} M.,
  2019, \mn@doi [\mnras] {10.1093/mnras/stz1703}, \href
  {https://ui.adsabs.harvard.edu/abs/2019MNRAS.487.5764T} {487, 5764}

\bibitem[\protect\citeauthoryear{{Tonnesen}, {Smith}, {Kollmeier}  \&
  {Cen}}{{Tonnesen} et~al.}{2017}]{tonnesen-2017}
{Tonnesen} S.,  {Smith} B.~D.,  {Kollmeier} J.~A.,   {Cen} R.,  2017, \mn@doi
  [\apj] {10.3847/1538-4357/aa7fb8}, \href
  {https://ui.adsabs.harvard.edu/abs/2017ApJ...845...47T} {845, 47}

\bibitem[\protect\citeauthoryear{{Tripp}, {Savage}  \& {Jenkins}}{{Tripp}
  et~al.}{2000}]{tripp-2001}
{Tripp} T.~M.,  {Savage} B.~D.,   {Jenkins} E.~B.,  2000, \mn@doi [\apjl]
  {10.1086/312644}, \href
  {https://ui.adsabs.harvard.edu/abs/2000ApJ...534L...1T} {534, L1}

\bibitem[\protect\citeauthoryear{{Tytler}, {Fan}  \& {Burles}}{{Tytler}
  et~al.}{1996}]{tytler-1996}
{Tytler} D.,  {Fan} X.-M.,   {Burles} S.,  1996, \mn@doi [Nature]
  {10.1038/381207a0}, \href
  {https://ui.adsabs.harvard.edu/abs/1996Natur.381..207T} {381, 207}

\bibitem[\protect\citeauthoryear{{Viel}, {Haehnelt}, {Bolton}, {Kim},
  {Puchwein}, {Nasir}  \& {Wakker}}{{Viel} et~al.}{2017}]{viel-2017}
{Viel} M.,  {Haehnelt} M.~G.,  {Bolton} J.~S.,  {Kim} T.-S.,  {Puchwein} E.,
  {Nasir} F.,   {Wakker} B.~P.,  2017, \mn@doi [\mnras]
  {10.1093/mnrasl/slx004}, \href
  {https://ui.adsabs.harvard.edu/abs/2017MNRAS.467L..86V} {467, L86}

\bibitem[\protect\citeauthoryear{{Vogelsberger} et~al.,}{{Vogelsberger}
  et~al.}{2014}]{vogelsberger-2014}
{Vogelsberger} M.,  et~al., 2014, \mn@doi [\mnras] {10.1093/mnras/stu1536},
  \href {https://ui.adsabs.harvard.edu/abs/2014MNRAS.444.1518V} {444, 1518}

\bibitem[\protect\citeauthoryear{{Wakker}, {Hernandez}, {French}, {Kim},
  {Oppenheimer}  \& {Savage}}{{Wakker} et~al.}{2015}]{wakker-2015}
{Wakker} B.~P.,  {Hernandez} A.~K.,  {French} D.~M.,  {Kim} T.-S.,
  {Oppenheimer} B.~D.,   {Savage} B.~D.,  2015, \mn@doi [\apj]
  {10.1088/0004-637X/814/1/40}, \href
  {https://ui.adsabs.harvard.edu/abs/2015ApJ...814...40W} {814, 40}

\bibitem[\protect\citeauthoryear{{Weinberg}, {Katz}  \& {Hernquist}}{{Weinberg}
  et~al.}{1998}]{weinberg-1998}
{Weinberg} D.~H.,  {Katz} N.,   {Hernquist} L.,  1998, in {Woodward} C.~E.,
  {Shull} J.~M.,   {Thronson} Harley~A. J.,  eds,  Astronomical Society of the
  Pacific Conference Series Vol. 148, Origins. p.~21 (\mn@eprint {arXiv}
  {astro-ph/9708213})

\bibitem[\protect\citeauthoryear{{Weinberger} et~al.,}{{Weinberger}
  et~al.}{2018}]{weinberger-2018}
{Weinberger} R.,  et~al., 2018, \mn@doi [\mnras] {10.1093/mnras/sty1733}, \href
  {https://ui.adsabs.harvard.edu/abs/2018MNRAS.479.4056W} {479, 4056}

\bibitem[\protect\citeauthoryear{{Whittam}, {Prescott}, {McAlpine}, {Jarvis}
  \& {Heywood}}{{Whittam} et~al.}{2018}]{whittam-2018}
{Whittam} I.~H.,  {Prescott} M.,  {McAlpine} K.,  {Jarvis} M.~J.,   {Heywood}
  I.,  2018, \mn@doi [Monthly Notices of the Royal Astronomical Society]
  {10.1093/mnras/sty1787}, \href
  {https://ui.adsabs.harvard.edu/abs/2018MNRAS.480..358W} {480, 358}

\bibitem[\protect\citeauthoryear{{de Graaff}, {Cai}, {Heymans}  \&
  {Peacock}}{{de Graaff} et~al.}{2019}]{degraaff-2019}
{de Graaff} A.,  {Cai} Y.-C.,  {Heymans} C.,   {Peacock} J.~A.,  2019, \mn@doi
  [\aap] {10.1051/0004-6361/201935159}, \href
  {https://ui.adsabs.harvard.edu/abs/2019A&A...624A..48D} {624, A48}

\makeatother
\end{thebibliography}

%\begin{thebibliography}{99}
%\end{thebibliography}

%%%%%%%%%%%%%%%%%%%%%%%%%%%%%%%%%%%%%%%%%%%%%%%%%%

%%%%%%%%%%%%%%%%%%%%%%%%%%%%%%%%%%%%%%%%%%%%%%%%%%

% Don't change these lines
\bsp	% typesetting comment
\label{lastpage}
\end{document}